\documentclass[twocolumn]{article}

\usepackage[utf8]{inputenc}
\usepackage[T1]{fontenc}
\usepackage{times}

\usepackage{wrapfig}

\usepackage{graphicx}
\usepackage{amsmath}
\usepackage{float}
\usepackage{textcomp}
\usepackage{authblk}

\usepackage{sectsty}
\usepackage[table]{xcolor}
\usepackage{colortbl}
\definecolor{Gray}{gray}{0.85}

\usepackage[left=1.1cm,right=1.1cm,top=2cm,bottom=1.7cm]{geometry}

\usepackage[labelsep=space]{caption}
\definecolor{light-gray}{gray}{0.8}
\definecolor{very-light-gray}{gray}{0.9}

\title{Modeling and hexahedral meshing of cerebral arterial networks from centerlines}
\author{Méghane Decroocq$^{1,2,4,5,6}$, Carole Frindel$^{1,5}$, Pierre Rougé$^{1}$, Makoto Ohta$^{4,5}$, Guillaume Lavoué$^{2,3}$}
\date{\bigbreak}
\affil{ $^1$ CREATIS, Université Lyon1, CNRS UMR5220, INSERM U1206, INSA-Lyon, 69621 Villeurbanne, France \\ 
$^2$ LIRIS, CNRS UMR 5205, F-69621, France \\
$^3$ Ecole Centrale de Lyon, France \\
$^4$ ELyTMaX IRL3757, CNRS, INSA Lyon, Centrale Lyon, Université Claude Bernard Lyon 1, Tohoku University, 980-8577, Sendai, Japan \\
$^5$Institute of Fluid Science, Tohoku University, 2-1-1, Katahira, Aoba-ku, Sendai, Miyagi 980-8577, Japan \\
$^6$ Graduate School of Biomedical Engineering, Tohoku University, 6-6 Aramaki-aza-aoba, Aoba-ku, Sendai, Miyagi 980-8579, Japan \\

Corresponding Author: \textit{carole.frindel@creatis.insa-lyon.fr}}

\begin{document}

\twocolumn[
  \begin{@twocolumnfalse}
    \maketitle
    \begin{abstract}
    
Computational fluid dynamics (CFD) simulation provides valuable information on blood flow from the vascular geometry. However, it requires extracting precise models of arteries from low-resolution medical images, which remains challenging. Centerline-based representation is widely used to model large vascular networks with small vessels, as it encodes both the geometric and topological information and facilitates manual editing. In this work, we propose an automatic method to generate a structured hexahedral mesh suitable for CFD directly from centerlines.

We adressed both the modeling and meshing tasks. We proposed a vessel model based on penalized splines to overcome the limitations inherent to the centerline representation, such as noise and sparsity. The bifurcations are reconstructed using a parametric model based on the anatomy that we extended to planar n-furcations. Finally, we developed a method to produce a volume mesh with structured, hexahedral, and flow-oriented cells from the proposed vascular network model. 

The proposed method offers better robustness to the common defects of centerlines and increases the mesh quality compared to state-of-the-art methods. As it relies on centerlines alone, it can be applied to edit the vascular model effortlessly to study the impact of vascular geometry and topology on hemodynamics. We demonstrate the efficiency of our method by entirely meshing a dataset of 60 cerebral vascular networks. 92\% of the vessels and 83\% of the bifurcations were meshed without defects needing manual intervention, despite the challenging aspect of the input data. The source code is released publicly.

\\ \\

    \end{abstract}
  \end{@twocolumnfalse}
]

\section{Introduction}
\label{sec:Introduction}

% Vascular disease and CFD
Cerebrovascular diseases, such as stroke, can cause severe disability or death \cite{ramos2018quality}. The relationship between the geometry of the vascular network and the onset and the outcome of the pathology is increasingly investigated in the literature. Computational fluid dynamics is a key tool for this type of study, as it provides information on the hemodynamics from the vessel geometry \cite{saqr2020does,sugiyama2016blood}. The main limitation of the use of CFD is the creation of the computational mesh. Indeed, numerical simulation requires a smooth, precise, and anatomically realistic mesh of the arterial wall to provide reliable results. In pathologies like ischemic stroke, the distribution of the vessels in the different vascular territories of the brain impacts the position and evolution of the lesion \cite{hodneland2019new}. It requires reconstructing large and complex cerebral arterial networks with small vessels whose radius is close to the image resolution, which remains very challenging. Besides, in the finite element and finite volume methods, the shape of the cells inside the volume affects the simulations. In particular, flow-oriented, structured hexahedral cells were shown to improve the stability of the simulation while lowering the computational cost \cite{vinchurkar2008evaluation,de2010patient,ghaffari2017}. Those results were confirmed in this work by running CFD experiments whose results are presented in Section \ref{sec:Applications}. Despite this, tetrahedral cells remain widely used due to their ability to automatically mesh any complex shape. The approaches investigated in the literature to address the meshing of vascular networks can be divided into two categories: the segmentation-based and the centerline-based methods. 

The segmentation of magnetic resonance angiography (MRA) images is a non-invasive way to access patient-specific vasculature. A lot of effort was put to develop efficient vessel-enhancing filters  \cite{jerman2015beyond,merveille2017curvilinear} and improve the segmentation methods. In particular, the rise of deep learning-based segmentation methods resulted in significant progress in vascular segmentation \cite{tetteh2020deepvesselnet, livne2019u}. However, the accuracy of the segmentation does not guarantee the accuracy of the mesh it entails (e.g. vessels merging due to the image resolution, disconnected vessels, bumps) nor its usability for numerical simulation. Besides, the vascular network is generally meshed with tetrahedral elements, and hexahedral re-meshing is not straightforward.

Following the tubularity assumption, vessels can be reduced to a centerline-radius description. Segmentation-based and centerline-based models complement each other, with centerline extraction being used as a pre-processing or post-processing of segmentation. Many methods to extract vessel skeletons from binary or raw images were proposed in the literature \cite{zhang2021confluent,he2020learning}. As opposed to image segmentation, centerline-based representation advantageously incorporates the network topology and enables manual extraction and editing. This simplified representation is more suitable for the construction of big databases of large vascular networks \cite{wright2013} or the creation of ideal models. It also offers more editing flexibility than segmentation-based meshes. As it encodes the vessel topology and orientation, it has a high potential for the creation of meshes with high-quality, flow-oriented cells. Nevertheless, the representation of vessels by centerlines lowers the geometric information content; depending on the extraction method, only a limited number of data points are used and noise can be introduced in the dataset. It causes inaccuracy in the shape of the vessels and the position and geometry of bifurcations. These limitations make it difficult to reconstruct a surface that matches the requirements of numerical simulation. In this article, we propose a method to overcome those limitations and create a high-quality, structured hexahedral mesh for CFD from centerlines only, opening the way to CFD in large cerebral arterial networks.

\section{Related work}
\label{sec:Relatedwork}

\subsection{Segmentation-based meshing}

Segmentation of medical images is the most common method to obtain patient-specific meshes for CFD. In recent years, deep learning-based models have led to significant advances in vascular segmentation. More specifically, convolutional neural networks (CNNs) have achieved very good performances \cite{jiang2018retinal,tetteh2020deepvesselnet}. The popular U-net architecture \cite{ronneberger2015u} has been successfully applied to the segmentation of intracranial vessels in \cite{quon2020deep} and \cite{livne2019u}. \cite{hilbert2020brave} proposed an extended U-net architecture using context aggregation and deep supervision for brain vessel segmentation. Besides, the attention mechanisms have been used to help the network to better learn global dependencies and increase the receptive field in \cite{mou2021cs2}, \cite{ni2020global} and \cite{li2021ta}.

For medical applications such as CFD, more than the segmentation itself, the smoothness and the topological accuracy of the mesh it entails are critical. However, in the literature, there was very little focus on the conversion of the segmented volumes to mesh. Recently, \cite{wickramasinghe2020voxel2mesh} and \cite{kong2021deep} introduced new neural network architectures to reconstruct 3D meshes directly from 3D image volumes. Despite those recent advances, the meshing largely relies on algorithms such as the marching cubes, followed by a smoothing step, to produce a surface mesh with tetrahedral elements \cite{watanabe2018hemodynamic,misaki2021inflow}. However, this type of segmentation-based meshes commonly suffers from geometrical and topological inaccuracies (e.g. merging or disconnected vessels, bulges, missing vessels) and requires burdensome manual post-processing \cite{glasser2015reconstruction}, as we demonstrate in Section \ref{subsec:Comparisonsegmentation}. Such problems are not correctly captured by the image-based metrics (e.g. DICE score) used to evaluate the segmentation methods. To overcome those challenges, the centerline representation of vascular networks has recently gained interest.  

Some recent segmentation approaches propose integrating the vessel centerline information to build more topology-oriented metrics. \cite{keshwani2020topnet} proposed to segment the vascular network from its skeleton by learning a connectivity metric between center-voxels. Besides, \cite{shit2021cldice} introduced a novel topology-preserving loss for the training of neural networks, which relies on the centerlines of the predicted segmentation. The information provided by the centerlines allowed the neural network to improve the topology correctness of the segmentations. In this context, we believe that the use of centerlines in the meshing process can offer many advantages for CFD applications.

\subsection{Centerline-based meshing}

In this part, we review the methods used to reconstruct the vascular surface from centerlines. The main issues to overcome in this task arise from the defects commonly observed in the vascular centerline extracted from medical images; local discontinuities causing a lack of information -especially at the bifurcation parts- and noise due to the voxelization. In this context, the smoothness of the vessel surface and the reconstruction of the bifurcations are important locks. The reconstruction methods can be divided into explicit methods, where a tetrahedral mesh of the surface is produced, and implicit methods where the surface is represented by implicit functions. Implicit methods employ radial basis functions \cite{hong2018accurate}, implicit extrusion surfaces \cite{hong2020high} or local implicit modeling \cite{kerrien2017blood} to reconstruct vascular networks from medical images or  centerlines \cite{abdellah2020interactive}. If they stand out by their ability to reconstruct complex branching topology, they do not allow as much editability and control on the final mesh (e.g. hexahedral meshing) as explicit methods.

In explicit methods, the vessel surface is created by sweeping along the centerlines. The quality of the reconstruction depends on the way centerline points are interpolated, usually with Bezier or spline functions \cite{guo2013mesh, kocinski2016centerline, ghaffari2017}. The proposed methods largely rely on interpolation and might be affected by the quality of the input centerline. The details of the methods employed and the evaluation of the accuracy of the estimation of coordinates and derivatives were not provided in previous studies. For the branching part, various bifurcation models were proposed. In the work of \cite{kocinski2016centerline} and \cite{ghaffari2015automatic}, the three branches of the bifurcation are modeled separately and joined at the bifurcation center. The junction is then blended to restore the continuity, by a subdivision scheme for \cite{kocinski2016centerline} and Bezier segments for \cite{ghaffari2015automatic}. This geometric model facilitates the creation of hexahedral meshes. However, the realism of the bifurcation shape depends on the accuracy of the position of the bifurcation center and the tangent of the branches, which is hard to estimate correctly from centerlines. In the same way, \cite{han2015design} and \cite{guo2013mesh} modeled bifurcations using three tubes connecting the inlet and outlet sections. Their method guarantees the smoothness of the model but results in unnatural-looking bifurcations. Finally, \cite{zakaria2008} proposed a parametric model where the bifurcations are represented by two merged tubes. It was validated concerning both the accuracy of the anatomy and the CFD simulations. As opposed to the bifurcation models presented above, it does not rely on the geometrical center of the bifurcation, but on a set of anatomical parameters (apex, apical sections, inlet, and outlet sections). However, the authors extracted the model parameters from a surface mesh, and they did not suggest a way to extract them from centerlines.

\subsection{Hexahedral meshing}

 For applications in CFD with the finite element or finite volume method, the inside of the surface mesh must be discretized into cells. The commonly used cell shapes include tetrahedral, prismatic, and hexahedral. Hexahedral meshes can be further divided into two categories; the structured meshes, where the neighborhood relationships between the cells are defined in the mesh structure (e.g. regular grid), and the unstructured mesh. In the case of blood vessels, structured and unstructured hexahedral meshing also allows for the creation of flow-oriented cells. Studies of the literature show that both the shape of the cells (tetrahedral, hexahedral) and the type of mesh (structured or unstructured) influence the cost and the stability of the numerical simulation.
 
\cite{vinchurkar2008evaluation}, \cite{de2010patient} and \cite{ghaffari2017} compared the performances of hexahedral and tetrahedral meshes for different models (airways, coronary tree, and cerebral arteries) and applications. Those studies demonstrated that hexahedral meshes in general, and more specifically structured hexahedral meshes, converge better for the same accuracy of the result. \cite{de2010patient} and \cite{ghaffari2017} reported that 6 times fewer cells (resp. 10 times) and 14 times (resp. 27 times) less computational time were required. Finally, \cite{vinchurkar2008evaluation} insisted on the importance of hexahedral flow-oriented cells for near-wall measurements (e.g. particle deposition, wall shear stress). The advantages of hexahedral cells are not limited to CFD; this type of mesh simplifies the boundary layer creation, bridges the gap between representation and physical simulation, and provides a basis for NURBS modeling \cite{zhang2007patient} and isogeometric analysis. Hexahedral meshing, and more specifically structured hexahedral meshing, is however limited by a far more complex generation process than standard tetrahedral meshes \cite{vinchurkar2008evaluation}.

	In the application to the arterial networks, the main challenge is the generation of the mesh at the bifurcations. In the literature, this task was addressed by a two-step approach; the bifurcations are first decomposed into three branches, and the hexahedral mesh is then created based on this decomposition. A variety of methods were proposed to obtain a robust branch decomposition. De Santis et al. introduced semi-automatic methods, ranging from the manual selection of the most relevant slices of the input surface mesh (\cite{de2010patient}), user-defined bifurcation coordinate system \cite{de2011patient}, to the generation and adjustment of a block-structure representation of the network \cite{deSantis2011}. Automatic methods are based on Voronoi diagram \cite{antiga2002geometric}, resolution of the Laplace's equation \cite{verma2005all}, random-walk algorithm \cite{xiong2013automated} or branching templates \cite{zhang2007patient}. A hexahedral mesh is created from the decomposition through various techniques; Copper scheme in the work of \cite{antiga2002geometric}, template grid sweeping for \cite{verma2005all}, \cite{zhang2007patient} and \cite{ghaffari2017}, template mesh fitting \cite{de2011patient}, projection and refinement of block-structures \cite{deSantis2011}, or Catmull-Clark subdivision \cite{xiong2013automated}. 

The main limitation of the application of the methods described above to our framework is that they rely on a triangular surface mesh as input for the extraction of a high-resolution skeleton, the branch decomposition, and the meshing. Only \cite{de2010patient} and \cite{ghaffari2017} proposed methods to create a hexahedral mesh directly from the centerlines, without resorting to triangular surface meshes. These methods are based on geometric models of bifurcations which directly provides a decomposition for hexahedral meshing and require high-sampled centerlines. Besides, only \cite{de2011patient} provided their code through the user-friendly interface PyFormex, which enables generating hexahedral meshes semi-automatically for a single bifurcation.

\subsection{Contributions}
\label{subsec:Contributions}

% Method description
In this work, our purpose is to meet the challenges arising from this state-of-the-art with a framework integrating a modeling and a meshing step. The shortcomings of segmentation-based meshing are addressed by developing a method based on centerlines. A parametric model is used to overcome the common defect of centerlines and reconstruct a realistic vessel surface. Finally, an original meshing algorithm is proposed to create high-quality structured hexahedral meshes for CFD simulations. Our main contributions are:

\begin{itemize}
\item{We introduced an original vessel model and its approximation algorithm based on penalized splines, which models both the spatial coordinates and the radius in a single function and offers good robustness to noise and low-sampling.}

\item{The vessel model is combined with an anatomical model of bifurcation \cite{zakaria2008}, to form a light parametric model of the entire vascular network. A method to extract the parameters of the bifurcation model directly from centerlines is proposed, and the model was generalized to planar n-furcations. If the bifurcation model itself is not new, the use of this type of anatomical bifurcation model - as opposed to geometric bifurcation models - to reconstruct a realistic vascular shape has not been investigated in previous studies.}

\item{We developed a parametric method to create a structured hexahedral volume mesh with flow-oriented cells from the vessel and bifurcation models proposed. It includes relaxation and smoothing steps to improve the quality of the cells without deforming the model shape. This method gives more control over the distribution and density of the cells than tetrahedral meshing.}

\item{The model and the mesh are stored in a graph structure which enables to easily and inexpensively edit the topology and geometry of the vascular networks.}

\item{The code is publicly released and can be downloaded at: https://github.com/megdec/vascularmd.}
\end{itemize}

As it is based on centerlines only, the proposed framework opens the way to numerical simulation in large cerebral vascular networks. It was evaluated qualitatively and quantitatively against other explicit and implicit centerline-based meshing methods, as well as segmentation-based meshing methods. Finally, several practical applications are presented, including the meshing of a large database of 60 large cerebral networks, pathology modeling, topology and geometry editing, and finally a CFD study comparing a healthy and stenotic middle carotid artery.

\section{Input data}
\label{sec:Inputdata}
		
	The input vessel centerlines we consider are composed of a set of data points with three spatial coordinates (x,y,z), radius value (r), and the connectivity between points. A data point can have several successors, for instance in bifurcations. A point with $n$ successors is called $n$-furcation. The centerlines are stored using the \textit{swc} format or VMTK format of \cite{izzo2018vascular}. In this work, we used two publicly available datasets. The Aneurisk database \cite{AneuriskWeb} provides 3D models of the main arteries of the circle of Willis for patients with an aneurysm. High-resolution centerlines were extracted from the surface meshes using the VMTK software. The BraVa database \cite{wright2013} provides the centerlines of the whole cerebral network for 60 patients. To create this dataset, the data points were manually placed by medical doctors on medical images using the ImageJ plugin Neurite Tracer \cite{longair2011simple}. The radius was computed automatically by the same tool. These centerlines have a lower spatial resolution and are prone to errors and uncertainty on the data point position and radius.

\section{Modeling}
\label{sec:Modeling}
	\subsection{Vessel model}
	\label{subsec:ModelVessels}
	
In this part, we focus on modeling vessels from centerline data; we will address the case of bifurcations in the next section. Different models of centerline have been proposed in the literature, based on the interpolation of data points by Bezier segments \cite{ghaffari2017}, regression splines \cite{kocinski2016centerline}, free knot regression splines or local polynomial smoothing \cite{sangalli2009efficient}. Only \cite{sangalli2009efficient} details the methods and provides a thorough study of the accuracy of the spatial coordinates and the derivatives. However, the accuracy of both the first and second derivatives is crucial, as the vessel curvature impacts the hemodynamics \cite{sangalli2009case}. Moreover, the meshing techniques often rely on the normals of the centerline \cite{kocinski2016centerline, ghaffari2017}. It is important to note that the proposed approximation methods \cite{sangalli2009efficient, kocinski2016centerline, ghaffari2017} focus on the spatial coordinates of the centerlines, excluding the radius. In this work, we propose a parametric model of vessels based on approximation by penalized splines. Our approximation method enables combining spatial coordinates and radius in a single function and is robust to the defects of the input data.
		 
		\subsubsection{Penalized splines}	
		
We want to approximate a set of $m$ points $\{ D_{0}, D_{1} ..., D_{m-1} \}$  with $4$ coordinates $(x, y, z, r)$, using a spline function $s$ defined as 

\begin{equation}
s(u) = \sum_{i = 0}^{n - 1} N_{i, p}(u) P_{i},
\label{Eq:spline}
\end{equation}

for $u \in [0,1]$, where $N_{i, p}$ is the $i$th basis spline function of order $p$ and $\{ P_{0}, P_{1}, ..., P_{n - 1} \}$ the $n$ control points of the spline. The part of the spline controlled by each control point is defined by a set of knots.

The main challenge in the approximation of noisy data is to find the optimal balance between the proximity of the curve to data points and the smoothness of the curve (i.e. the accuracy of the derivatives). There are two main approaches to control the smoothness of a spline function. The first is to adjust the number of control points: a low number of control points will result in a smoother curve. In this case, the position of the knots can be optimized like in \cite{sangalli2009efficient}. The other approach is to use a large number of control points and a uniform knot vector and to constrain the smoothness by a penalty on the second derivatives \cite{craven1978smoothing, eilers1996flexible}. For reasons further detailed below, we judged the second approach more suitable for our task. The vessels are modeled with penalized splines, as introduced by \cite{eilers1996flexible}. The control points are optimized using the two-term cost function given in Equation \ref{Eq:cost-function}. The first term accounts for the closeness to the data points and the second term accounts for the smoothness of the approximation spline. The parameter $\lambda$ gives the balance between both terms. The cost function is defined as

\small
\begin{equation}
f(P_{0},..., P_{n - 1}) = \sum_{k = 0}^{m-1}{|D_{k} - s(t_{k})|^{2}} + \lambda \sum_{j = 2}^{n}{(P_{j}-2 P_{j-1}+P_{j-2})^2},
\label{Eq:cost-function}
\end{equation}
\normalsize

where $t$ is a time parametrization vector that associates each data point to a position on the spline. 

		\subsubsection{Approximation strategy}

Centerline data are composed of spatial coordinates $(x, y, z)$ and radii $r$, two variables of different scales that might show different levels of noise. For this reason, we propose to approximate them separately. The choice of penalized splines allows us to dissociate $\lambda$ values for the position and the radius in a two-step approximation algorithm. With this approach, the spatial and radius coordinates are modeled with a single spline. 

For the approximation, we use a uniform knot vector and a parametrization obtained by the chord-length method. We set the number of control points so that the non-penalized approximation curve (i.e. produced by solving equation \ref{Eq:cost-function} with $\lambda=0$) has a root mean square distance from the original data lower than a given value, which is set in this work to $10^{-1}$ for spatial coordinates and $10^{-3}$ for the radius.

We first solve the linear system arising from equation \ref{Eq:cost-function} for the spatial coordinates $(x,y,z)$ of the centerline data points. The system is written as 

\begin{equation} 
P_{(x,y,z)}  = (N^{T}N + \lambda_{s} \Delta)^{-1}N^{T}D_{(x,y,z)},
\label{Eq:solution-spatial}
\end{equation}

where $N$ is the matrix of representation of the basis spline functions, and $\Delta$ is the matrix representation of the difference operator that appears in the second term of the cost function \ref{Eq:cost-function}.
We set the optimal value for $\lambda_{s}$ by minimizing the Akaike criterion $AIC_{2}$, as detailed in Section \ref{subsec:Vesselmodelevaluation}. A comparison study with other optimization criteria, given in Supplementary Materials, led to the choice of this criterion. 
 
The linear system is then solved for the data $(t, r)$ where $t$ is the time parametrization of each data point and $r$ their radius value:

 \begin{equation} 
P_{(t,r)}  = (N^{T}N + \lambda_{r} \Delta)^{-1}N^{T}D_{(t,r)}.
\label{Eq:solution-radius}
\end{equation}

\begin{figure}[!h]
\centering
\includegraphics[width=0.5\textwidth]{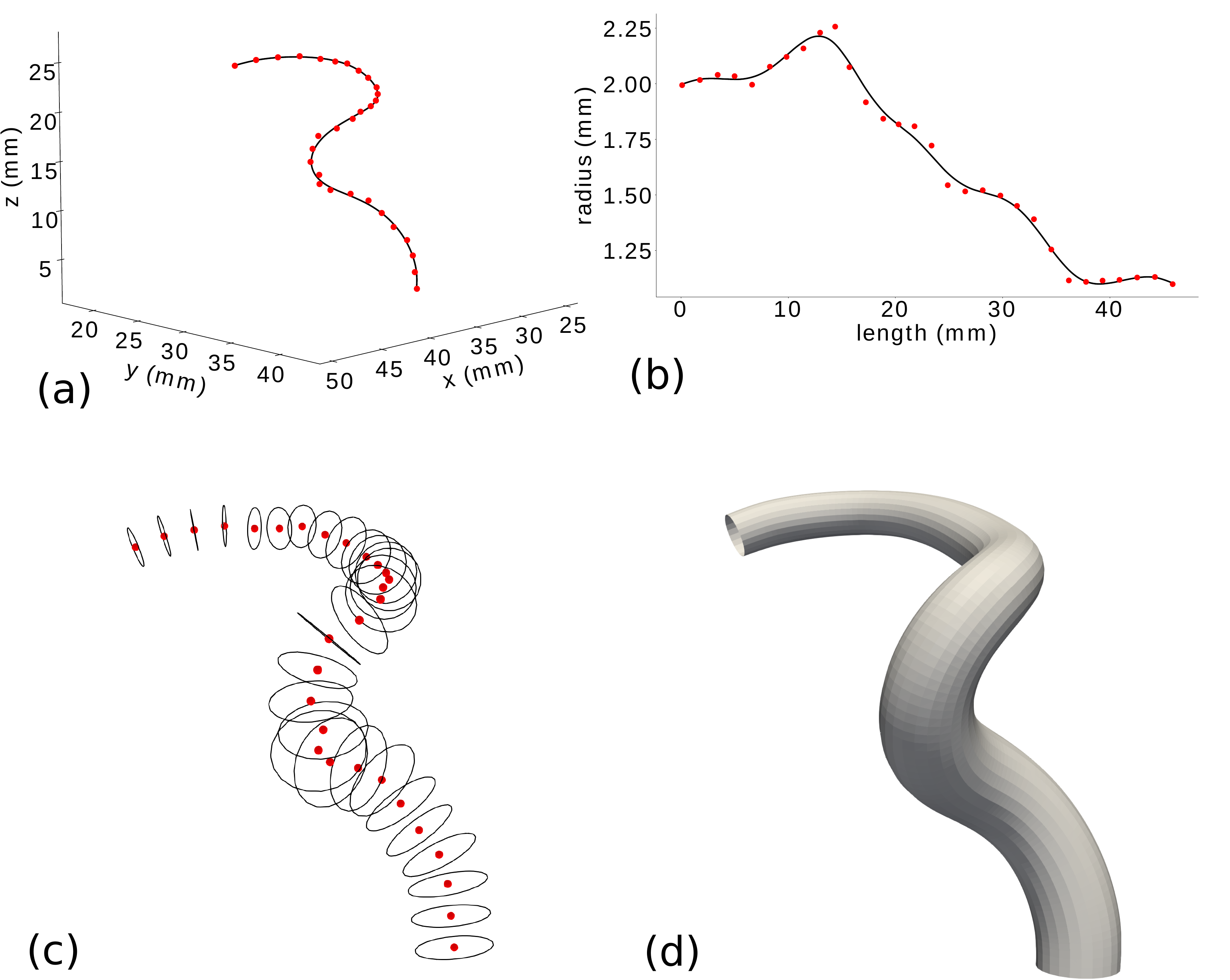}
\caption{Approximation of a noisy centerline with the proposed method. Figures (a) and (b) show respectively the approximation of the spatial coordinates and the radius. Figure (c) shows the input centerline data as red dots with a radius represented by black circles. Figure (d) represents the vessel surface defined by the approximating spline.}
\label{Img:approximation}
\end{figure}

We also select the value of $\lambda_{r}$ by minimizing the Akaike criterion on the time/radius data. The spatial coordinates and radius of the optimized control points are then concatenated to form the 4-coordinates control points of the final spline. Figure \ref{Img:approximation} illustrates this two-part approximation scheme. In Section \ref{subsec:Vesselmodelevaluation}, we compare the proposed approximation method with other conventional approximation methods regarding the robustness to noise and low sampling of the data points.

	\subsection{Bifurcation model}
	\label{subsec:ModelBifurcations}
	
In this part, we focus on modeling bifurcations.
	
		\subsubsection{Zakaria's model}
		
\begin{figure}[!h]
\centering
\includegraphics[width=0.25\textwidth]{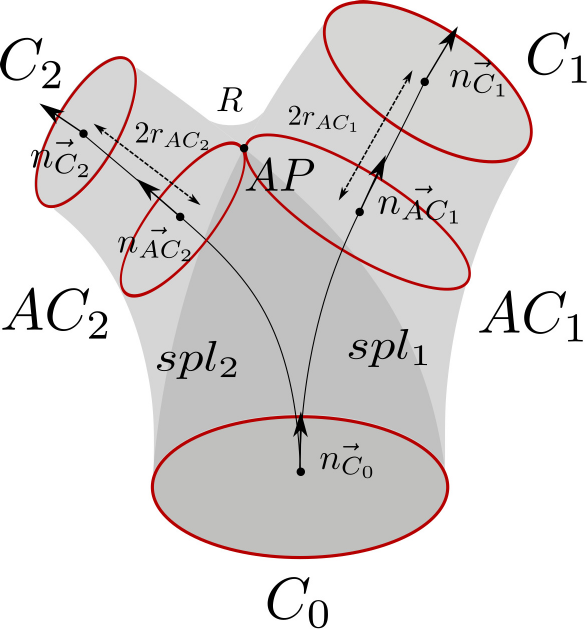}
\caption{The five cross-sections bifurcation model introduced by \cite{zakaria2008}.}
\label{Img:bifurcation_model}
\end{figure}

	 \cite{zakaria2008} proposed a parametric model for non-planar bifurcations. This model was validated regarding the anatomy and the numerical simulation of blood flow and showed a good agreement with real cerebral bifurcations. It requires only a few parameters and is well-suited for the reconstruction of bifurcations from a few data points. In this model, bifurcations are modeled as the combination of two tubes sharing the same inlet. The tubes are defined by three cross-sections; a shared inlet cross-section $C_{0}$, separate apical cross-sections $AC_{1}$, $AC_{2}$ and outlet cross-sections $C_{1}$ and $C_{2}$. The apical cross-sections $AC_{1, 2}$ are located at the apex point $AP$ of the bifurcation. The outlet sections $C_{1, 2}$ are cut one diameter from the apex. In total, five cross-sections and their normals are required to build the model. Each circular cross-section $C$ is represented by the three spatial coordinates of its center $P_{c}$, the radius $r_{c}$ and the normal vector $\vec{n_{c}}$.
The centerline of each tube is defined by a spline function $spl_{1,2}$. The first segment of the centerline connects the inlet section $C_{0}$ to the apical section, and the second connects the apical section to the outlet sections. The tangent of the centerline segments matches the normal of the cross-sections they connect. The radius along the segments evolves linearly between $r_{C_{0}}$, $r_{AC_{1,2}}$ and $r_{AC_{1,2}}$, $r_{C_{1,2}}$. The bifurcation model is illustrated in Figure \ref{Img:bifurcation_model}. The unphysiological sharp angle produced between tubes at the apex is rounded by a segment of constant radius of curvature $R$.

		\subsubsection{Parameter estimation}
		
\begin{figure}[!h]
\centering
\includegraphics[width=0.5\textwidth]{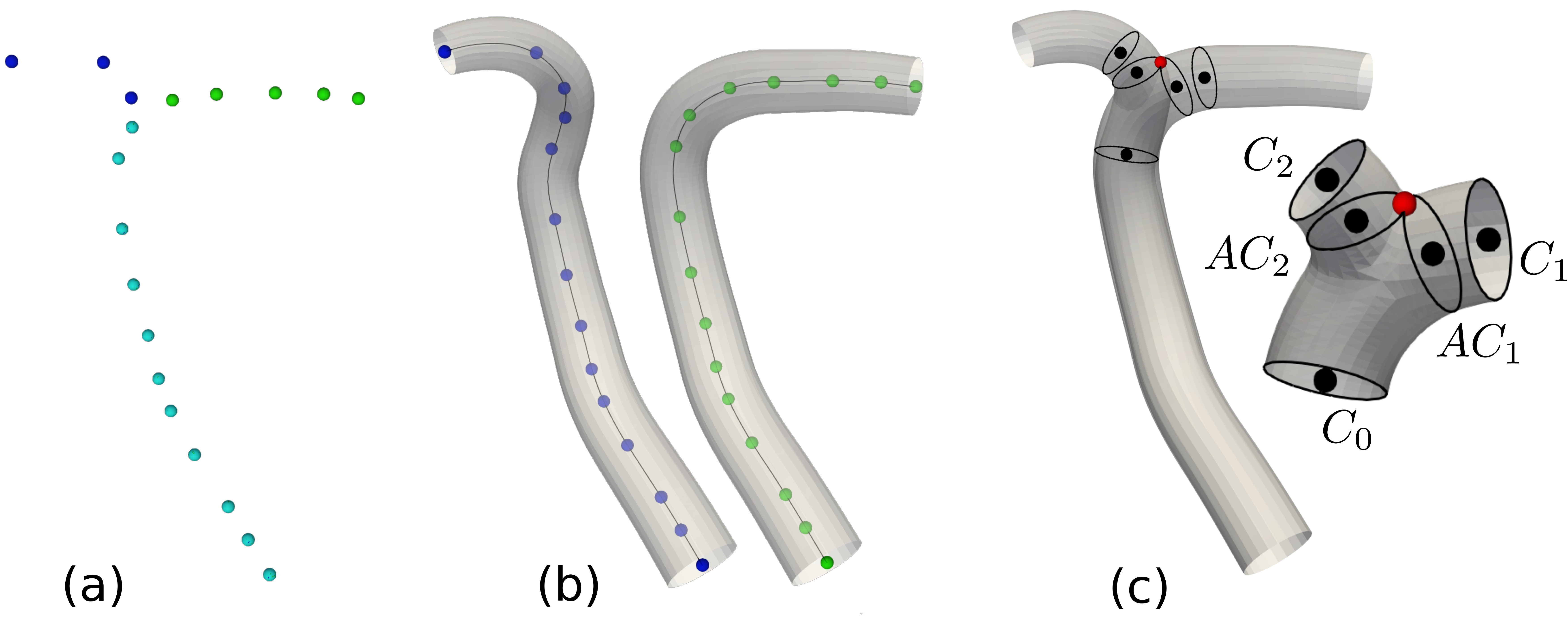}
\caption{Pipeline of the bifurcation parameter estimation. (a) shows the inlet and outlet data points, (b) the independent vessel models, and (c) the parameter extraction and resulting bifurcation.}
\label{Img:parameter_estimation}
\end{figure}

	To apply this bifurcation model to our framework, we introduce an algorithm to estimate the parameters of the bifurcations directly from the input centerline data. For all the bifurcations in the centerline data, the inlet data points (in light blue in Figure \ref{Img:parameter_estimation} (a)) are concatenated with each of the outlet data points (resp. in deep blue and green in Figure \ref{Img:parameter_estimation} (a)) to form two separate vessel centerlines going through the bifurcation, as shown in Figure \ref{Img:parameter_estimation} (b). The two vessels based on these centerlines are modeled independently by splines using the approximation strategy presented in section \ref{subsec:ModelVessels}. We compute the apex $AP$ of the bifurcation as the point where the surface of the two vessel models first intersect (red dot on Figure \ref{Img:parameter_estimation} (c)). $AP$ is then projected on the model splines $spl_{1}$ and $spl_{2}$. The tangent and position of the obtained projection points then define the normal and the center of the apical cross-sections $AC_{1}$ and $AC{2}$. The outlet sections $C_{1}$ and $C_{2}$ are computed from the evaluation of the spline where the length from the apex projection point is twice the radius of the apical section. 
		
		\subsubsection{Tangent continuity}
		
	 The full vascular network model is created by assembling the vessels and bifurcations models presented in the previous sections. To preserve the continuity of the different parts of the network, the inlet and outlet tangents of the vessels must match the normal of the inlet and outlet cross-sections of the bifurcations they connect. For this, we introduce an additional constraint on the end-points and tangents in the resolution of the approximation equations \ref{Eq:solution-spatial} and \ref{Eq:solution-radius} used to model the vessels. \cite{piegl2000least} proposed a least-square spline approximation with arbitrary end derivatives. To limit the influence of the end constraints on the approximation of data points, we propose a weaker constraint that fixes the end tangent while the derivative is free. We consider a spline $s$ as defined by Equation \ref{Eq:spline}. Because we work with clamped curves, $s(0) = P_{0}$ and $s(1) = P_{n-1}$. Moreover,  $s'(0)$ (respectively $s'(1)$) is in the same direction as vector $P_{1}$ - $P_{0}$ (respectively $P_{n-2}$ - $P_{n-1}$). If we note $S_{0}$ and $S_{n-1}$ the fixed end-points and $T_{0}$ and $T_{n-1}$ the fixed end tangents, the following new conditions are applied to the system \ref{Eq:solution-spatial}:
	
\begin{equation}
\begin{cases}
P_{0} = S_{0} \\
P_{n-1} = S_{n-1}\\
P_{1} = P_{0} + \alpha T_{0}\\
P_{n-1} = P_{n-2} + \beta T_{n-1},\\
\end{cases}
\end{equation}

where $\alpha$ and $\beta$, the end tangent magnitude, are additional parameters to optimize. Those constraints guarantee the $G_{1}$ continuity of the network. We give the details of the system resolution in Supplementary Materials.

\section{Structured hexahedral meshing}
\label{sec:Structuredhexahedralmeshing}

In this section, we present the meshing algorithm developed to produce a hexahedral mesh with flow-oriented cells from the parametric model described in Section \ref{sec:Modeling}.

	\subsection{Bifurcation surface mesh}
	\label{subsec:Bifurcations}
		\subsubsection{Decomposition}
		
\begin{figure}[!h]
\centering
\includegraphics[width=0.40\textwidth]{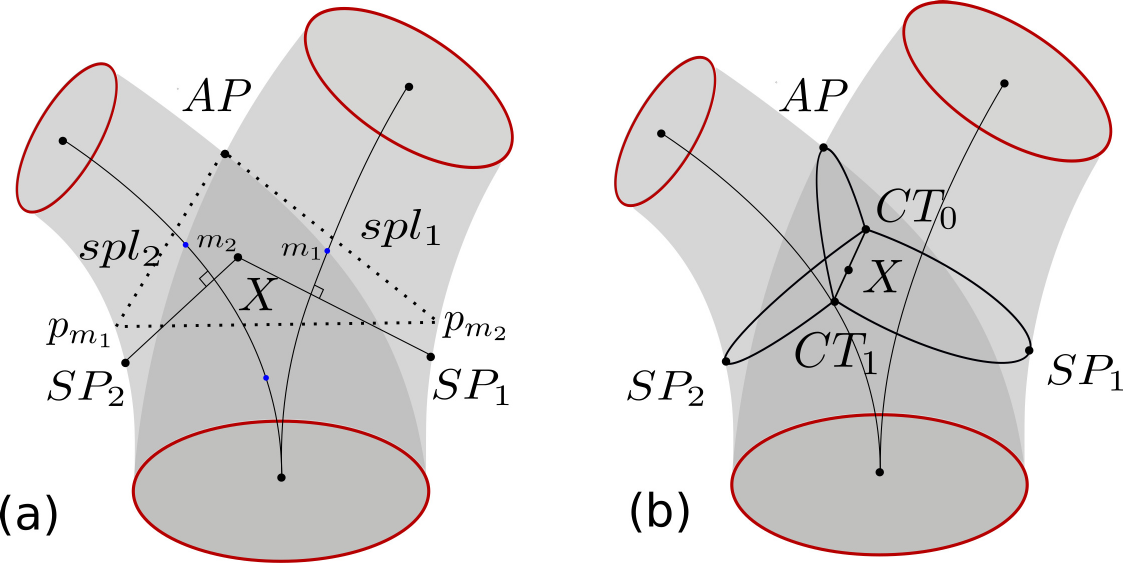}
\caption{Geometric decomposition of the bifurcation model. In (b), the end cross-sections are represented in red, and the separations planes are in black.}
\label{Img:bifurcation_decomposition}
\end{figure}

	The bifurcations are the most challenging parts to mesh with hexahedral elements. We propose a decomposition method to split the bifurcation into three geometrical branches; one inlet branch and two outlet branches. This method allows retrieving the meshing advantages of geometrical bifurcation models of other methods of the literature while keeping the realism of the anatomical bifurcation model used in this work. Figure \ref{Img:bifurcation_decomposition} (b) gives an example of branch splitting using three separation planes. Antiga et al. \cite{antiga2004robust} proposed a bifurcation decomposition scheme based on the Voronoi diagram of a surface mesh. Based on this work, we introduced a decomposition method based on the geometrical information of our bifurcation model. 

In this decomposition method, we define three separation planes using a set of five points; the apex point $AP$, which is already a parameter of the model, two center points $CT_{0}$ and $CT_{1}$ and two separation points $SP_{1}$, and $SP_{2}$. As illustrated in Figure \ref{Img:bifurcation_decomposition} (a), we first define the geometric center of the bifurcation $X$, as the barycenter of $AP$, $p_{m_{1}}$ and $p_{m_{2}}$, where $p_{m_{1, 2}}$ are the projection of the key points $m_{1,2}$ located at the intersection of one centerline with the surface of the other vessel. The separation points $SP_{1, 2}$ are obtained by projecting $X$ on the surface in the opposite direction from $AP$.

Finally, the position of center points $CT_{0}$ and $CT_{1}$ is obtained by projecting $X$ on the surface of the vessels. The direction of projection is normal to the plane defined by the three points $AP$, $SP_{1}$, and $SP_{2}$. The separation points $AP$, $SP_{1}$ and $SP_{2}$ are finally connected to the center points $CT_{0}$, $CT_{1}$ by arcs, which delineate a geometrical frontier between the branches of the bifurcation (see Figure \ref{Img:bifurcation_decomposition} (b)), providing the desired branch decomposition. This decomposition method enables us to handle large radius differences between the daughter vessels, as the barycenter $X$ is naturally closer to the vessel with the smallest radius,  relaxing the angles between the separation planes. 

		\subsubsection{Initial mesh}
		
\begin{figure}[!h]
\centering
\includegraphics[width=0.25\textwidth]{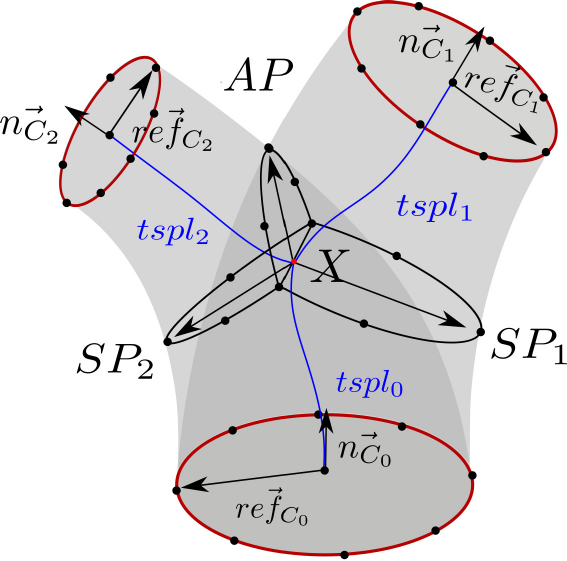}
\caption{Computation of the nodes (black dots) of the end cross-sections and the separation planes for $N=8$ and splines $tspl_{1,2,3}$.}
\label{Img:bifurcation_meshing}
\end{figure}

\begin{figure}[!h]
\centering
\includegraphics[width=0.4\textwidth]{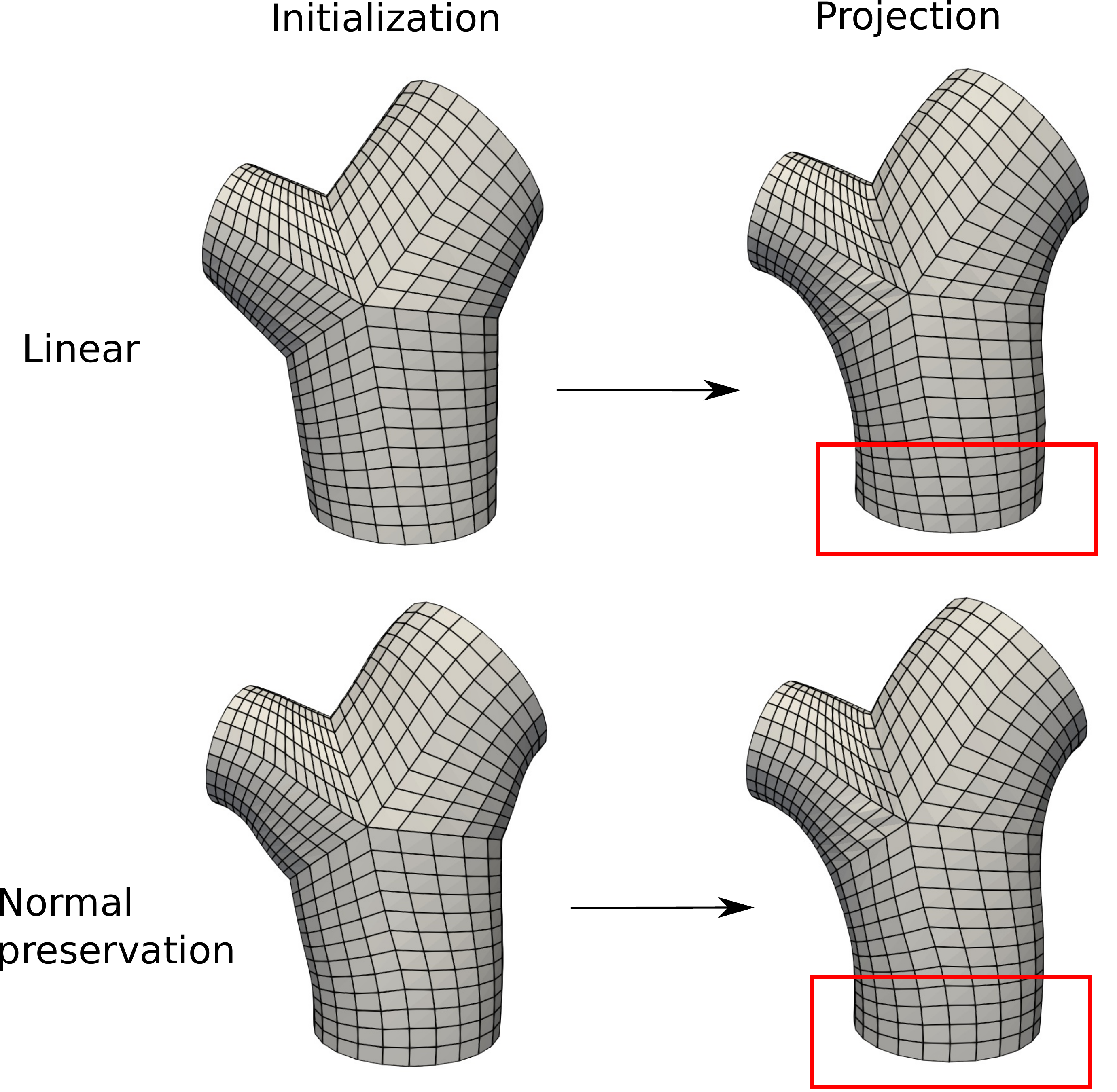}
\caption{Initial surface mesh and mesh after projection for the two types of initialization considered. The red squares emphasize the impact of the two types of initialization on the final mesh.}
\label{Img:surface_trajectories}
\end{figure}

In this step, we create the surface mesh of the bifurcation using the separation planes defined in the previous section. First, we compute an initial mesh grid connecting the end cross-sections to the separation planes by a set of successive cross-sections, as illustrated in Figure \ref{Img:surface_trajectories}. The cross-sections composing the grid have a number $N$ of nodes, with $N$ a multiple of $4$. Figure \ref{Img:bifurcation_meshing} illustrates the initial mesh creation process. We first compute the position of the $N$ nodes of the end cross-sections $C_{0}$, $C_{1}$ and $C_{2}$. For this, for each cross-section $C$, we define a normalized reference vector $\vec{ref_{C}}$ that minimizes the rotation with the separation points $SP_{1,2}$. The nodes are placed on the outline of the cross-section with evenly spaced angles starting by $\vec{ref_{C}}$ and rotated counterclockwise. Then, the nodes of the separation planes are positioned with equally sampled angles along the arcs connecting the separation point $AP$, $SP_{1}$ and $SP_{2}$ to both center points $CT_{0}$ and $CT_{1}$. The nodes of the end cross-sections and the nodes of the separation half-sections are connected to form the initial mesh grid, as shown in Figure \ref{Img:surface_trajectories}.

To create this grid, we define an initialization of the 3D trajectory connecting two nodes, as shown in the left column of Figure \ref{Img:surface_trajectories}. This initialization is an approximation used to control the topology and geometry of the final mesh grid, but it does not necessarily lie on the exact surface of the bifurcation at this stage. The initial trajectories are evenly sampled with $n$ nodes, where $n$ determines the number of cross-sections to compute along a given branch. $n$ is proportional to the radius of the branch, by a coefficient $d$ that can be adjusted to obtain the intended density of faces in the mesh. 

The nodes are then projected radially to the surface of the two vessels, as illustrated in the right column of Figure \ref{Img:surface_trajectories}. The direction of the projection is important to maintain the quality of the faces of the initial grid after projection. Ideally, the nodes of the initial trajectory must be displaced only radially from the center of the branch vessel. However, the shape splines $spl_{1}$ and $spl_{2}$ do not constitute a good approximation of the centerline of the three geometric branches. For this reason, we create another set of splines $tspl_{1,2,3}$ connecting the center of each end section to the center $X$ of the bifurcation, represented in blue in Figure \ref{Img:bifurcation_meshing}. We project the nodes to the surface according to the normal of this new set of splines.

\hspace{2cm}

% Initialization trajectories
The properties of the resulting mesh depend on the initial trajectory approximation. Figure \ref{Img:surface_trajectories} illustrates the meshes obtained after projection considering two types of initialization. The first row shows the simple case where we connect the nodes linearly.  In the second row, connection trajectories are computed so that the normal of the end sections is preserved in the output surface mesh. If both approximations allow preserving the topology of the grid and the quality of the faces after projection, the initial trajectories with normal preservation are closer to the actual surface of the bifurcation, causing less displacement of the nodes during projection. Besides, the preservation of the normal of the end sections facilitates the inclusion of the bifurcations in larger arterial networks as the bifurcation mesh can be smoothly extended to downstream vessels. In the rest of this work, we use the normal preserving initialization.

		\subsubsection{Relaxation}
		
\begin{figure}[!h]
\centering
\includegraphics[width=0.5\textwidth]{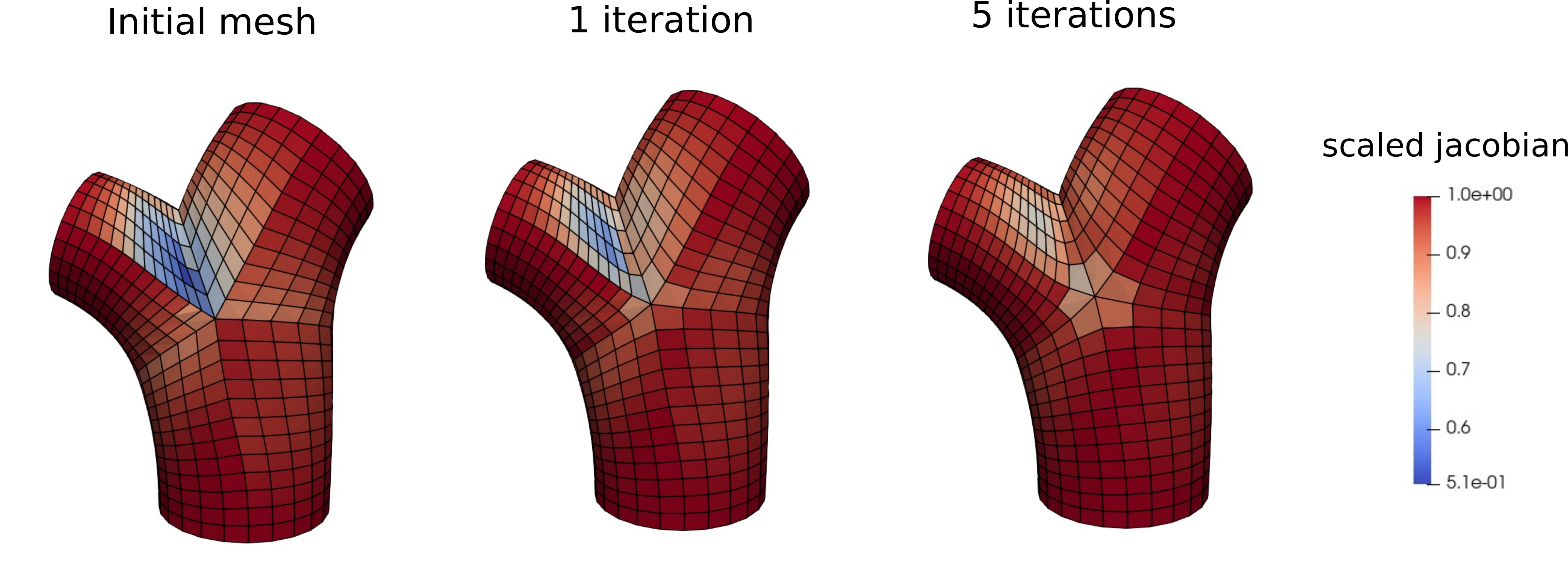}
\caption{Original bifurcation mesh and mesh after 1 and 5 relaxation iterations. The quality of the cells is measured by the scaled Jacobian metric, ranging between $-1$ (poor quality) and $1$ (high quality).}
\label{Img:resample}
\end{figure}

		The projection step results in an uneven sampling of the nodes along the bifurcation mesh that can produce faces with heterogeneous sizes or important skewness. We also observe a rupture of continuity at the separation between branches. We correct those unwanted features by a relaxation of the nodes of the surface mesh. If mesh smoothing methods are a simple way to reduce the skewness of faces, it triggers important deformations of the general shape of the model. To avoid deformations, \cite{vidal2015low} proposed to combine smoothing with a back projection. Following this approach, we first apply an iteration of Laplacian smooth (relaxation factor of $0.8$) to the bifurcation mesh. Then, we project the nodes back to the original surface. To prevent cross-sections from intersecting, the projection is made radially from the center of the cross-section. This process is repeated until the relaxation is satisfying. Figure \ref{Img:resample} displays a bifurcation mesh after 1 and 5 relaxation iterations. The faces are colored according to their geometric quality, measured by the scaled Jacobian metric. We observe that while the shape of the model is preserved, the quality of the faces near the separation planes is improved, and the grid now smoothly crosses the separation planes. Based on the average quality of the faces, we estimated that 5 relaxation iterations give the best results.
		
\hspace{2cm}

		\subsubsection{Apex smoothing}
		
		The last step of the bifurcation meshing is to smooth the apical region. Indeed, the bifurcation model presents an unwanted sharp angle where the two vessels merge. However, the curvature in the apex region impacts the pressure and velocity fields obtained by numerical simulation, as shown by \cite{haljasmaa2001effect}. Conventional mesh smoothing methods (e.g. Laplacian, Taubin smoothing) can produce smooth meshes with high-quality faces. However, as global methods, they struggle to generate important local deformations, resulting in unwanted modifications of the downstream and upstream branches. \cite{zakaria2008} proposed to smooth the apex region by projecting the nodes on a sphere of a given radius, rolling on the surface. This method is local, but it is computationally expensive and might not preserve the quality of the cells in the case of hexahedral meshes. Taking advantage of the topology of our hexahedral surface mesh, we propose a method to reduce this complex 3-dimensional problem to a 2-dimensional problem. 

\begin{figure}[h!]
\centering
\includegraphics[width=0.5\textwidth]{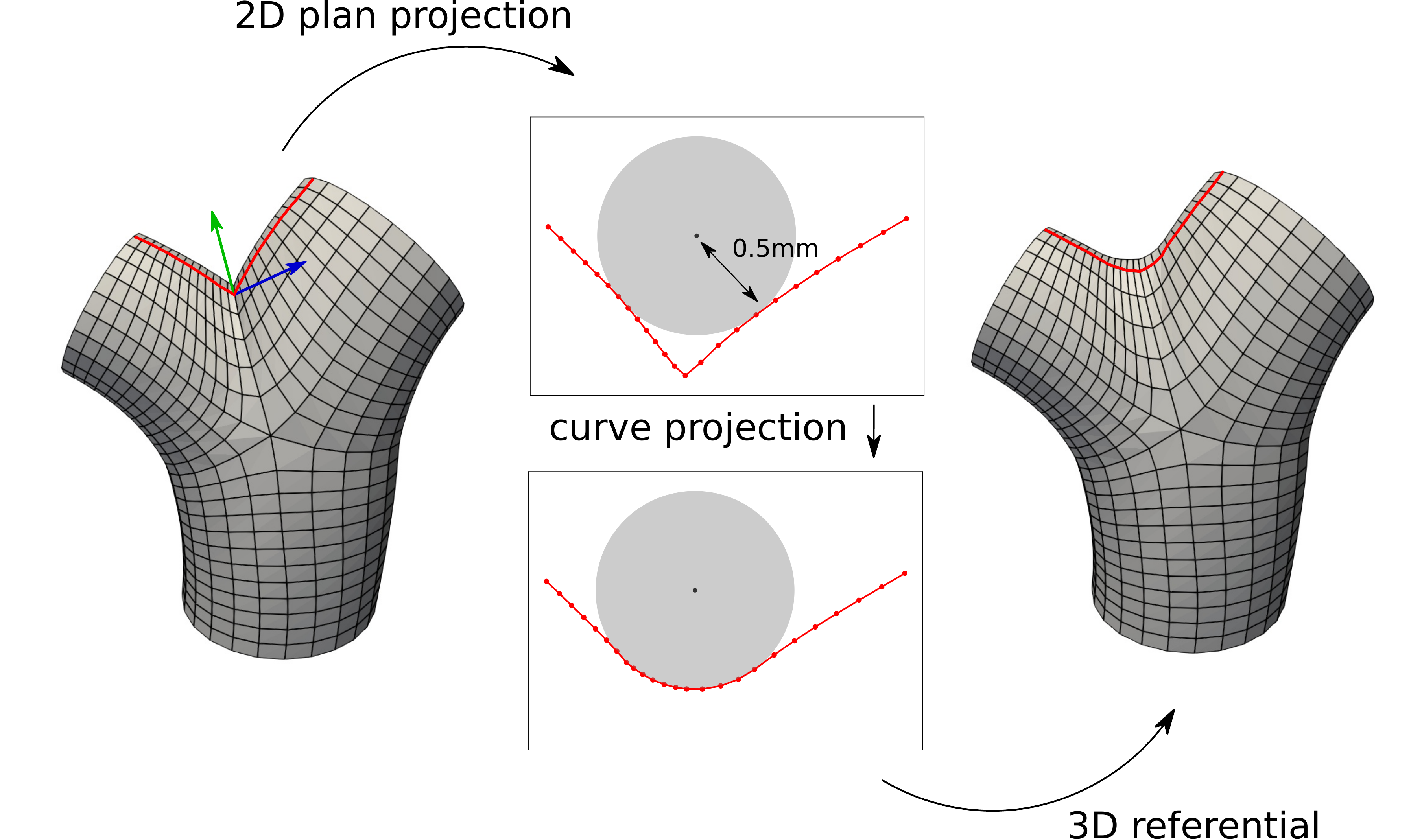}
\caption{Illustration of the apex smoothing pipeline.}
\label{Img:smoothing-pipeline}
\end{figure}

% Smoothing with inscribed circle
Figure \ref{Img:smoothing-pipeline} illustrates the proposed smoothing method. The curves connecting two nodes of the end cross-sections of the bifurcation are extracted from the 3D mesh (in red in Figure \ref{Img:smoothing-pipeline}). They are then projected on the 2D plane defined by the normal of the mesh at the separation point and the normal of the separation plane (resp. green and blue arrows on Figure \ref{Img:smoothing-pipeline}). A circle whose radius corresponds to the desired apex radius of curvature is rolled along those 2D curves. The position where the circle intersects the curve is computed analytically. We project the nodes onto the circle outline while preserving their original sampling. Finally, the new coordinates of the nodes are projected back on the original 3D referential to form the output surface mesh.

\begin{figure}[h!]
\centering
\includegraphics[width=0.5\textwidth]{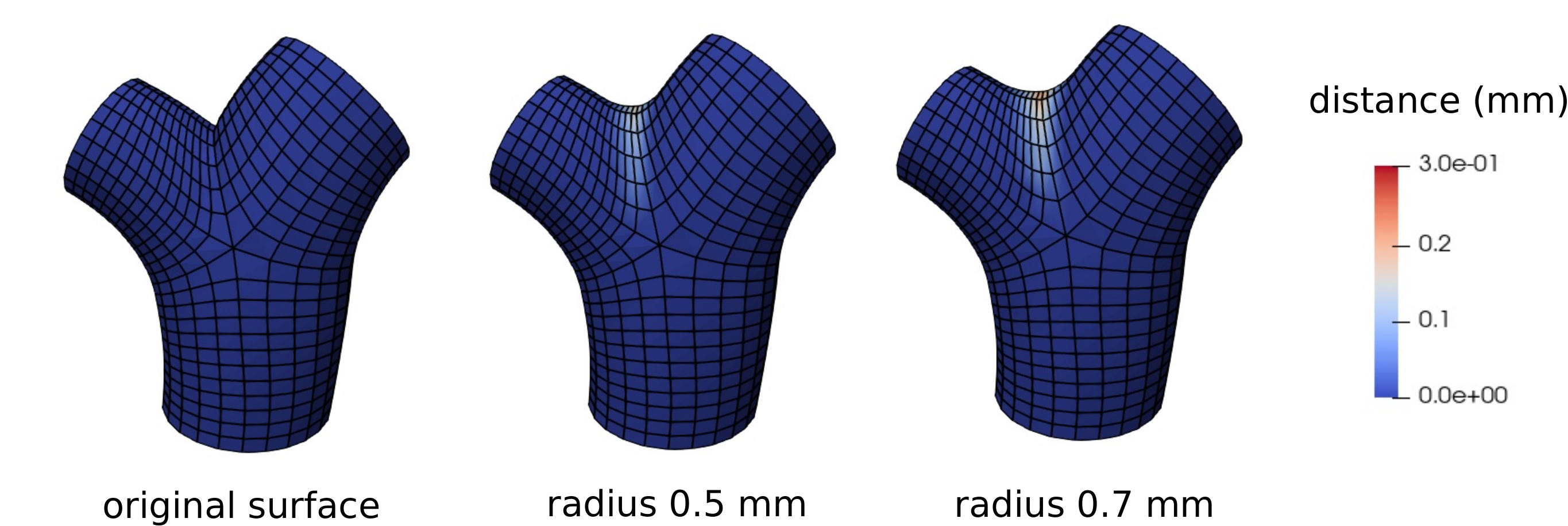}
\caption{Apex smoothing with different radius of curvature values. The colormap encodes the local distance to the original mesh, on the left.}
\label{Img:smoothing-example}
\end{figure}

The described smoothing method gives more control over the direction of projection and the sampling of the projected nodes, preserving the quality of the faces. As shown in Figure \ref{Img:smoothing-example}, the smoothing is local and does not affect the shape of the bifurcation outside of the apical region.

		\subsubsection{Planar n-furcations}
		
\begin{figure}[h!]
\centering
\includegraphics[width=0.5\textwidth]{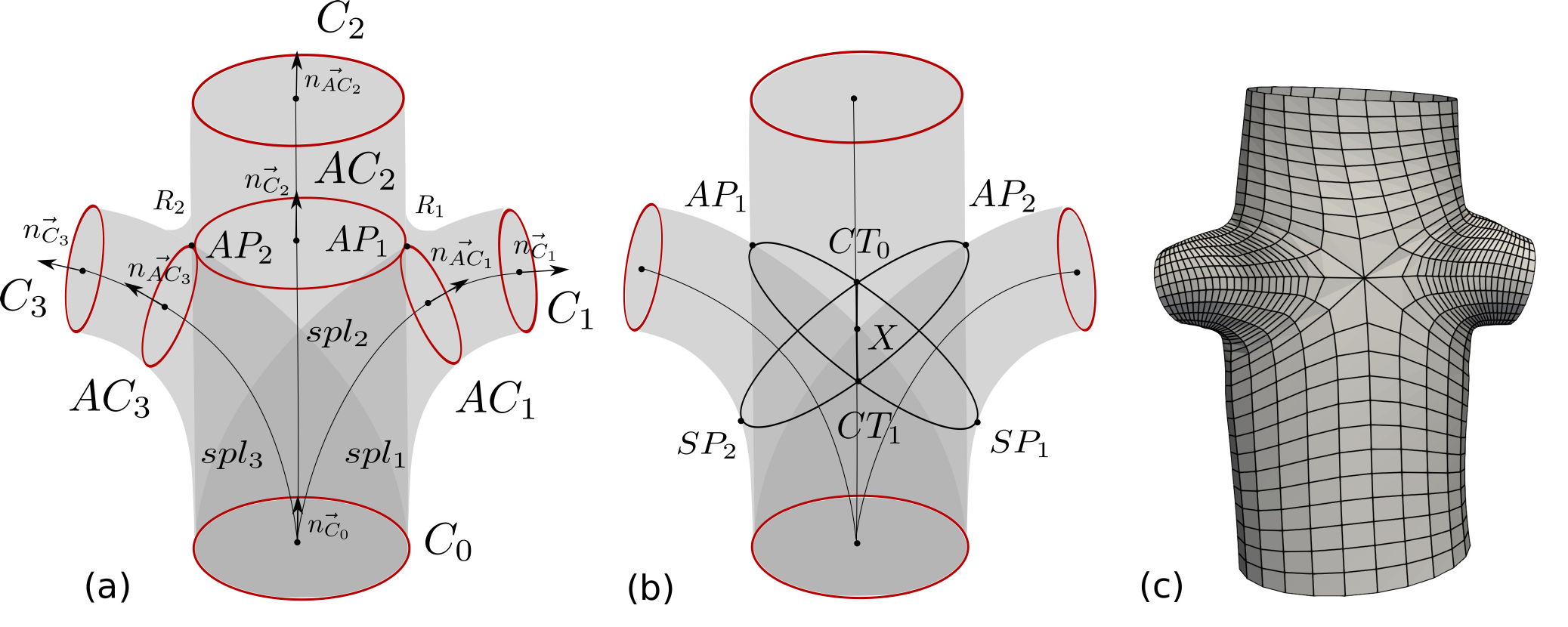}
\caption{(a) and (b) respectively illustrate the parametric model and the branch decomposition scheme for a trifurcation. (c) shows an example of trifurcation mesh.}
\label{Img:planar_trifurcation_decomposition}
\end{figure}

		If the cerebral arterial network is composed of a majority of bifurcations, multifurcations may also be present (e.g. trifurcations are frequently found on the basilar artery). To address this requirement, we generalized the model of \cite{zakaria2008} to planar n-furcations. The generalized n-furcation model is built with $n-1$ splines, $2n + 1$ cross-sections and $n-1$ apex points, as illustrated for the case $n=3$ in Figure \ref{Img:planar_trifurcation_decomposition} (a). We adjusted the decomposition scheme presented in section \ref{subsec:Bifurcations} to compute $n+1$ separation planes, as in Figure \ref{Img:planar_trifurcation_decomposition} (b). Figure \ref{Img:planar_trifurcation_decomposition} (c) shows an example of a planar trifurcation mesh obtained with this generalization. 

	\subsection{Vessel surface mesh}
	\label{subsec:Vessels}

		We adjusted the meshing method proposed by \cite{ghaffari2017} to the parametric model proposed in section \ref{subsec:ModelVessels} to obtain a quadrangular surface mesh of the vessels. The spline representing the vessel is evaluated at a set of time values equally sampled to set the center position and radius of the cross-sections along the vessel (i.e. the longitudinal resolution of the mesh). The density of cross-sections (number of cross-sections per $mm$) is proportional to the mean radius of the vessel, with a proportional coefficient $d$ that can be set by the user. From each center position, $N$ nodes are radially projected on the model surface to form a circular cross-section. The projection vector is swept along the centerline so that the twisting between cross-sections is minimized. The successive cross-sections are connected to form the mesh faces. In the case of vessels connecting one bifurcation to another, an extra rotation is smoothly applied to the cross-sections along the vessels so that the last cross-section of the vessel is aligned with the first cross-section of the next bifurcation. The surface meshes of the bifurcations and vessels are combined to create a quadrangular surface mesh of the vascular network.
		
	\subsection{Volume mesh}
		
\begin{figure}[!h]
\centering
\includegraphics[width=0.35\textwidth]{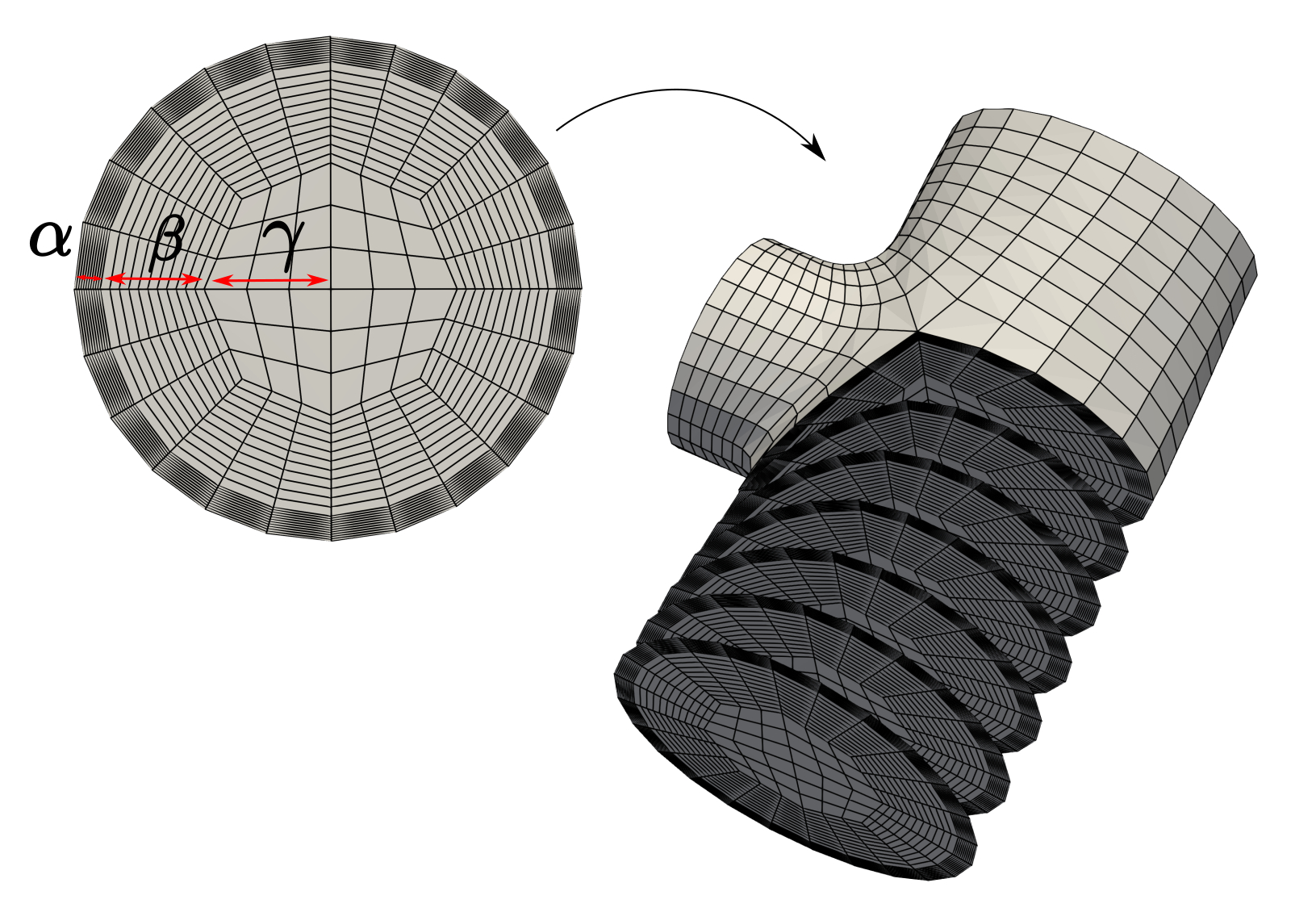}
\caption{Illustration of the O-grid pattern and volume meshing method.}
\label{Img:volume_mesh}
\end{figure}

	The quadrangular surface mesh is converted into a hexahedral volume mesh following the method proposed by \cite{ghaffari2017}. The cross-sections of the surface mesh are filled with a structured O-grid pattern. This pattern is composed of three different areas; the boundary layers, the intermediary layers, and the central block. The relative size $\alpha$, $\beta$, $\gamma$ of the areas, the number $N_{\alpha}$ of boundary layers and the number $N_{\beta}$ of intermediary layers can be adjusted. The separation planes of the n-furcations are handled by combining $n+1$ halves grids. The successive O-grid patterns are connected to form the hexahedral cells of the volume mesh, as shown in Figure \ref{Img:volume_mesh}.

	\subsection{Data encoding}
	
	In the proposed framework, we store the information of the model and mesh in graph structures. We create four graphs, corresponding to the different steps of the process; data, topology, model, and mesh. The geometric and topological information (e.g. centerline data points, model parameters, mesh nodes) are stored in the nodes and edges of the graph. This storage method facilitates data manipulation and editing as it allows the use of graph theory-based algorithms such as depth-first search or neighbor identification. The advantages of this data structure are further demonstrated in Section \ref{subsec:Topologyediting}.
	
\section{Results}
\label{sec:Results}

In this section, we evaluate the modeling and the meshing msethods proposed. We first assess the robustness and accuracy of the proposed vessel modeling method using a synthetic dataset of distorted centerlines. Then, we compare our meshing pipeline quantitatively and qualitatively with two concurrent state-of-the-art methods: deep learning-based segmentation and implicit meshing. We provide additional performance indicators regarding cell quality and computational time. 

	\subsection{Vessel model evaluation}
	\label{subsec:Vesselmodelevaluation}
	
	In this part, we evaluate the robustness of the approximation method presented in section \ref{subsec:Vessels} to defect commonly observed in centerlines.
	 
	\subsubsection{Validation dataset}
	
	 For this evaluation, we built a dataset of ground truth vessel centerlines. We selected four surface meshes of cerebral arteries from the Aneurisk database. From these meshes, we extracted a high-sampled, low-noise centerline of a single vessel using the VMTK software. The vessel to extract was selected so that it does not include pathologies, but goes through bifurcations, where we generally observe high curvature and important radius change. We created the ground-truth models from these centerlines by fitting data points with a spline $s$. To guarantee the accuracy of this model, we manually added control points to the spline and visually checked the accuracy of the fitting regarding spatial coordinates, radius, and first derivatives, until the approximation was judged satisfying. Ground truth models and their creation process are illustrated in Supplementary Material, Section 1.1. 
	 
These ground truth centerlines were then distorted to mimic defects commonly observed in centerlines; low sampling and noise. We applied spatial noise and radius noise separately, as they might differ in level. To generate spatial noise, the data points are displaced from their original position, with a distance randomly picked from a zero-centered Gaussian distribution with standard deviation $\sigma_{spatial}$. The direction of displacement is the normal of the ground truth spline $s$, so that it does not lead to unwanted radius noise. Radius noise is generated from a Gaussian distribution centered on the original radius value, of standard deviation $\sigma_{radius}$. The standard deviation values $\sigma_{spatial}$ and $\sigma_{radius}$ are proportional to the radius, as indicated in Table \ref{Tab:parameters_validation}, to keep similar levels of noise between large and small vessels. Finally, we modified the sampling of the centerlines by removing data points to reach a target density.

\begin{table}[!h]
\centering
\caption{Parameters used for the distortion of the ground truth centerlines}
\begin{tabular}{|cccccc|}
\hline
density $(mm^{-1})$ & $2$ & $4$ & $10$ & $16$ & $20$ \\
\hline
$\sigma_{radius} (mm)$  & $0.01r$ & $0.05r$ & $0.1r$ & $0.3r$ & $0.5r$ \\
\hline
$\sigma_{spatial} (mm)$  & $0.01r$ & $0.05r$ & $0.1r$ & $0.3r$ & $0.5r$ \\
\hline
\end{tabular}
\label{Tab:parameters_validation}
\end{table}

For each density value in Table \ref{Tab:parameters_validation}, we applied ten combinations of noise parameters; we added radius noise according to the values of $\sigma_{radius}$ given in Table \ref{Tab:parameters_validation}, as we set the spatial noise to $0$, and reversely. We repeated each combination three times to account for the stochastic effect. In total, we get $150$ distorted centerlines for each ground-truth vessel, bringing the total number of distorted centerlines in the dataset to $600$.

	  \subsubsection{Approximation methods}

To demonstrate the robustness and the accuracy of the approximation strategy used to reconstruct the surface of the vessels presented in Section \ref{subsec:ModelVessels}, we compared it to other methods of the literature \cite{kocinski2016centerline, ghaffari2017}. For this, we implemented four commonly used spline-based approximation methods with incremental complexity to emphasize the contributions of the proposed method.
\begin{itemize}

\item{\textbf{Global Non-Penalized} (GNP): The centerline data points are interpolated, without smoothness penalty in the cost function (Equation \ref{Eq:cost-function} with $\lambda = 0$). We fix the number of control points to match the RMSE threshold given in Section \ref{subsec:ModelVessels}.} 

\item{\textbf{Global Non-Penalized with Akaike criterion} (GNP-AIC): As in GNP, we do not apply any smoothness penalty during the fitting. The smoothness of the spline is controlled by adjusting the number of control points. The optimal number of control points minimizes the Akaike information criterion (\cite{akaike1998information}) $AIC_{1}$: 

\begin{equation}
AIC_{1} = m \times log(SSE) + 8(n + p).
\label{Eq:AIC-number-point}
\end{equation}

where $m$ is the number of data points, $p$ is the degree of the spline, $n$ is the number of control points and SSE is the sum squared error from the data points, including their four coordinates.}

\item{\textbf{Global Penalized with Akaike criterion} (GP-AIC): This approach corresponds to the original approximation by penalized splines described in \cite{eilers1996flexible}. It uses a smoothness penalty controlled by a global parameter $\lambda \neq 0$ as in Equation \ref{Eq:cost-function}.}

\item{\textbf{Spatial coordinates and Radius Penalized with Akaike criterion} (SRP-AIC): This approach corresponds to the approximation strategy that we propose in this work. Spatial and radius dimensions are approximated separately with two smoothing parameters $\lambda_{s}$ and $\lambda_{r}$. The comparison of our strategy with GP-AIC allows us to evaluate the contribution of treating the spatial and radius coordinates individually.}
\end{itemize}

 In methods GP-AIC and SRP-AIC, the criterion used to optimize the $\lambda$ values is another formulation of the Akaike information criterion ($AIC_{2}$), adapted to penalized splines, as proposed by \cite{eilers1996flexible}:  

\begin{equation}
AIC_{2} = m \times log(SSE/m) + 2tr,
\label{Eq:AIC}
\end{equation}
where  $tr$ is the trace of the matrix $H = N(N^{t}N + \lambda \Delta)^{-1}N^{t}$.

Unlike $AIC_{1}$, it is not employed to choose an optimal number of control points but to select the optimal value for the smoothing parameter $\lambda$.  

	 \subsubsection{Quality metrics}
	 
\begin{figure}[h!]
\centering
\includegraphics[width=0.28\textwidth]{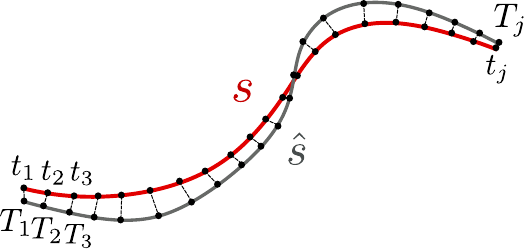}
\caption{Matching time parameters by minimum distance projection from $s$ onto $\hat{s}$}
\label{Img:matching_times}
\end{figure}

	 We proposed six quality metrics to evaluate the approximation strategies presented in the previous paragraph. We measured the distance between the ground truth spline $s$ and the approximation spline $\hat{s}$ by projection. As illustrated in Figure \ref{Img:matching_times}, two matched sets of time parameters are built. The spline $s$ is equally sampled with a time vector $t$, that we project on $\hat{s}$ according to the minimum distance to find the matched time vector $T$. The values associated with $t$ and $T$ are then compared.

We use the root mean squared error (RMSE) to measure the closeness of the approximation spline to the ground truth spline. The spatial coordinates and the radius values are treated separately in the evaluation. We note $\mathrm{RMSE_{radius}}$ (respectively $\mathrm{RMSE_{spatial}}$) the root mean squared error of the radius (respectively the spatial coordinates). To have a robust comparison between the curves, the projection is computed in both ways (from $s$ to $\hat{s}$ and from $\hat{s}$ to $s$) and the final RMSE value is the average of the RMSE yielded by both projections. 

The metrics $\mathrm{RMSEder_{spatial}}$ and $\mathrm{RMSEder_{radius}}$ measure the accuracy of the first derivatives of the model. As curvature is an important metric for blood flow study, its accuracy is assessed by the metric $\mathrm{RMSEcurv}$. Finally, the vessel length affects the delay of blood arrival between the inlet and the outlet of the vascular tree in numerical simulations. Therefore, we also considered the length difference $\mathrm{L_{diff}}$ in our evaluation.

\subsubsection{Results}

\begin{figure}[!h]
\centering
\includegraphics[width=0.5\textwidth]{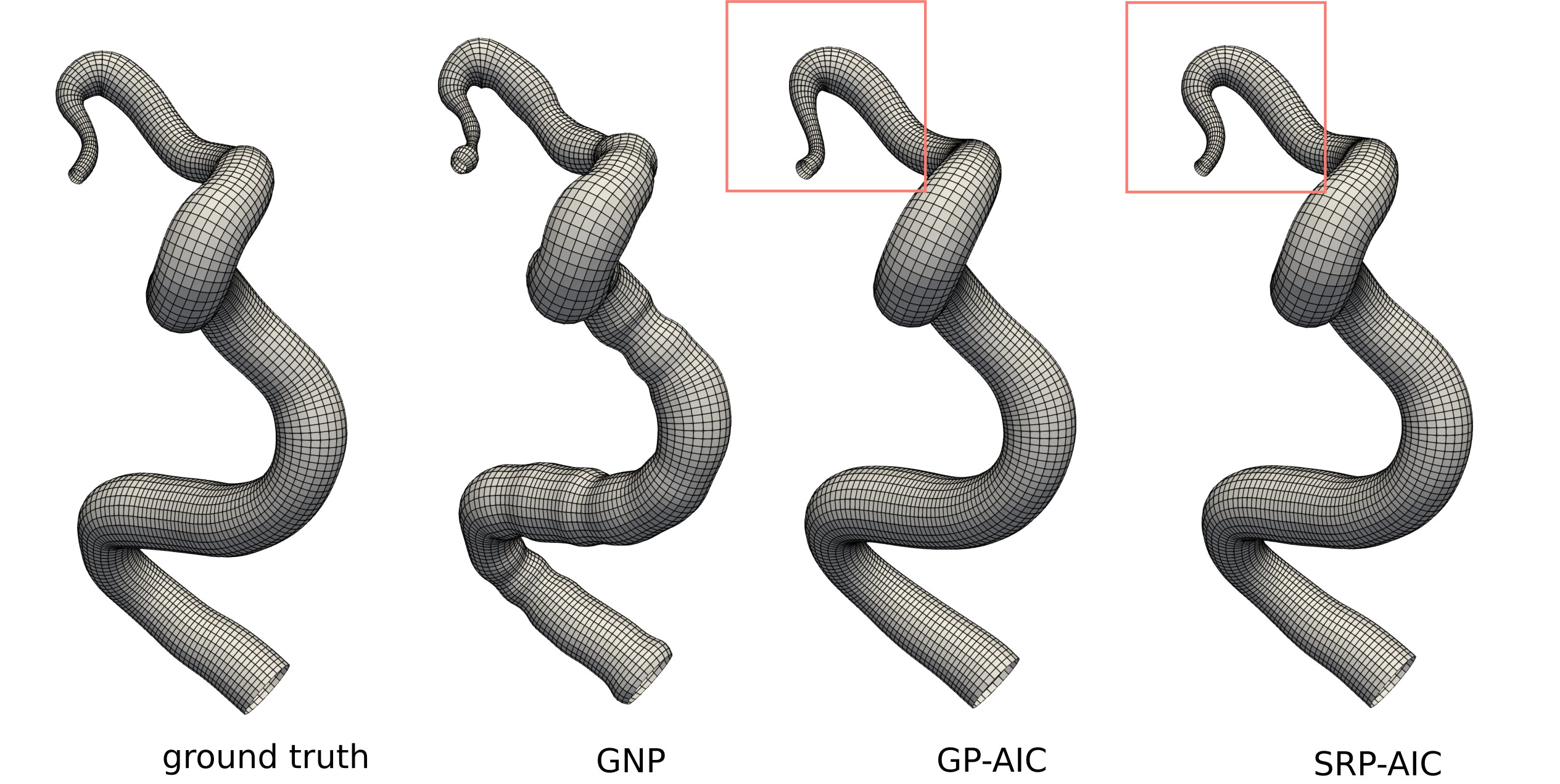}
\caption{Mesh resulting from the approximation of distorted data ($density = 1 mm^{-1}$, $\sigma_{radius} = 0.1$) by three of the methods compared in section \ref{subsec:Vesselmodelevaluation}.}
\label{Img:mesh_models}
\end{figure}

	As the spatial and radius distortions are not comparable in nature and magnitude, the evaluation results are presented in two different tables. Table \ref{Tab:validation-radius} (respectively Table \ref{Tab:validation-spatial}) shows the mean values of the six quality criteria for the four methods after radius noise (respectively spatial noise) addition. As expected, the non-penalized model (GNP) is sensible to the added noise and performs poorly for all radius-related metrics. In Figure \ref{Img:mesh_models}, the radius estimation error is visible on the vessel produced by this method. In the same way, the spatial-related metrics are impacted when spatial noise is added (Table \ref{Tab:validation-spatial}). In addition, a tendency to overfit the data is observed in Table \ref{Tab:validation-radius}, causing a very high spatial error. The overfitting and noise problems are partially solved by optimizing the number of control points with the method GNP-AIC. However, this approach still yields a poor approximation of the derivatives: as the number of control points is lower, the space between data points might not be correctly interpolated, which particularly impacts the curvature values.

The penalized approximations GP-AIC and SRP-AIC drastically improved the estimation of the derivatives and curvature. The advantage of SRP-AIC over GP-AIC is demonstrated both in the result tables \ref{Tab:validation-radius} and \ref{Tab:validation-spatial} and in Figure \ref{Img:mesh_models}. The global smoothing penalty used in GP-AIC forces a trade-off between the radius and spatial accuracy. In Figure \ref{Img:mesh_models}, if the radius of the vessel produced by GP-AIP is close to the ground truth, the trajectory of the centerline was impacted. On the other hand, SRP-AIC produced both accurate radius and trajectory. In conclusion, the proposed approximation method shows good robustness to the defects of the input data and models accurately the vessel trajectory and radius in a single function. More results are provided in Supplementary Material, Section 1.2.

\begin{table}[!h]
\caption{Overall evaluation of the approximation methods: mean values of the quality criteria for all the centerlines distorted by radius noise addition. The cells in gray correspond to the lowest error for each metric.}
\label{Tab:validation-radius}
\centering
\addtolength{\tabcolsep}{-1.5pt}
\begin{tabular}{|ccccc|}
  \hline
 & GNP & GNP-AIC & GP-AIC & SRP-AIC\\ 
  \hline
$\mathrm{RMSE_{spatial}}$ & 8.462 & 0.034 & 0.053 & \cellcolor{gray!25}0.029 \\ 
$\mathrm{RMSE_{radius}}$ & 17.523 & 0.095 & \cellcolor{gray!25}0.042 & 0.043 \\ 
$\mathrm{RMSEder_{spatial}}$ & 0.218 & 0.118 & 0.042 & \cellcolor{gray!25}0.009 \\ 
$\mathrm{RMSEder_{radius}}$ & 0.391 & 0.214 & \cellcolor{gray!25}0.032 & \cellcolor{gray!25}0.032 \\ 
$\mathrm{RMSEcurv}$ & 1919.428 & 190.531 & 0.060 & \cellcolor{gray!25}0.035 \\ 
$\mathrm{L_{diff}}$ & 718.906 & 0.057 & 0.207 & \cellcolor{gray!25}0.004 \\ 
   \hline
\end{tabular}
\end{table}

\begin{table}[!h]
\caption{Overall evaluation of the approximation methods: mean values of the quality criteria for all the centerlines distorted by spatial noise addition. The cells in gray correspond to the lowest error for each metric.}
\label{Tab:validation-spatial}
\centering
\addtolength{\tabcolsep}{-1pt}
\begin{tabular}{|ccccc|}
  \hline
 & GNP & GNP-AIC & GP-AIC & SRP-AIC\\ 
  \hline
$\mathrm{RMSE_{spatial}}$ & 0.511 & 0.152 & 0.099 & \cellcolor{gray!25}0.096 \\ 
$\mathrm{RMSE_{radius}}$ & 0.008 & 0.009 & 0.018 & \cellcolor{gray!25}0.007 \\ 
$\mathrm{RMSEder_{spatial}}$ & 0.314 & 0.343 & \cellcolor{gray!25}0.075 & 0.076 \\ 
$\mathrm{RMSEder_{radius}}$ & 0.015 & 0.019 & 0.021 & \cellcolor{gray!25}0.013 \\
$\mathrm{RMSEcurv}$ & 1.524 & 2.362 & \cellcolor{gray!25}0.085 & 0.091 \\ 
$\mathrm{L_{diff}}$ & 50.180 & 15.071 & 0.252 & \cellcolor{gray!25}0.207 \\ 
   \hline
\end{tabular}
\end{table}

	\subsection{Comparison with state-of-the-art methods}
	\label{subsec:Comparisonsegmentation}
	
 In this section, meshes obtained with our method are visually and quantitatively compared to meshes produced by state-of-the-art deep learning-based segmentation methods \cite{tetteh2020deepvesselnet,livne2019u}, as well as a recent implicit centerline-based meshing method \cite{abdellah2020interactive}.
	
		\subsubsection{Comparison pipeline}
		\label{subsec:comparison-pipeline}
		
\begin{figure}[!h]
\centering
\includegraphics[width=0.5\textwidth]{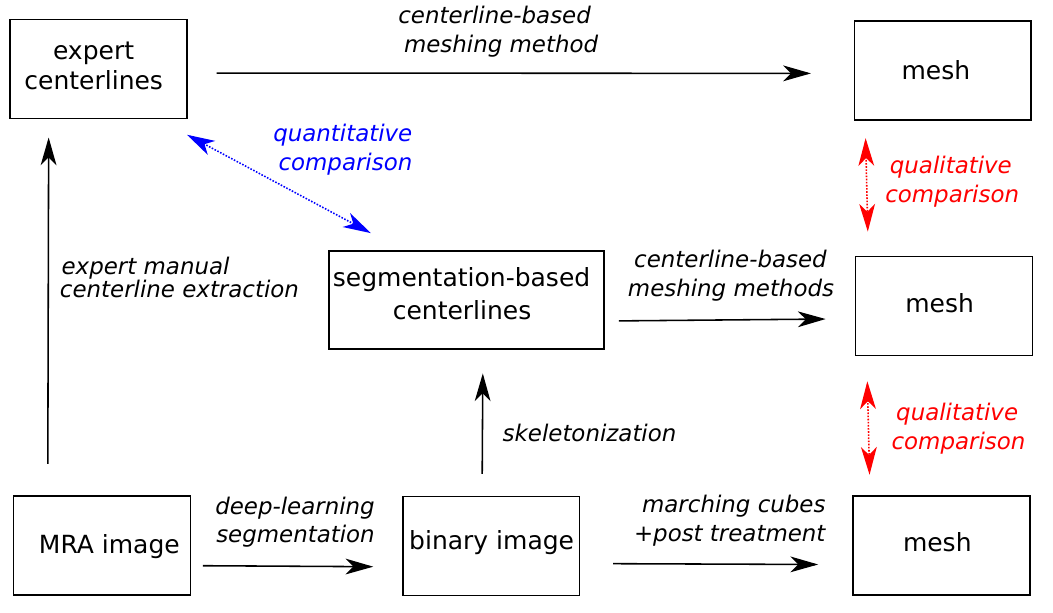}
\caption{Segmentation-based meshing and centerline-based meshing comparison pipeline.}
\label{Img:comparison-pipeline}
\end{figure}

% Segmentation methods and databases
	Centerlines can be extracted directly from the grayscale image or a segmented image. In this way, centerline-based meshing can be used either as a substitute or a complement to segmentation. In this comparison study, we investigated both approaches, as illustrated in Figure \ref{Img:comparison-pipeline}. We consider the centerlines manually extracted by medical experts \cite{wright2013} as reference for this study and refer to them as \textit{expert centerlines.}
	
	For the first part of our comparison pipeline (in blue in Figure \ref{Img:comparison-pipeline}), we segment the vascular network from MRA images using state-of-the-art segmentation methods \cite{tetteh2020deepvesselnet,livne2019u}. From this segmentation, we extract centerlines, referred to hereafter as \textit{segmentation-based centerlines}. In Section \ref{subsec:quantitative-eval}, we compare quantitatively these segmentation-based centerlines to the expert centerlines, as a way to evaluate the topological and geometrical correctness of the segmentation-based mesh.

In the second part of the comparison pipeline (in red in Figure \ref{Img:comparison-pipeline}), we use the extracted centerlines as input for the centerline-based meshing methods. To match the input requirements of the proposed method, the centerlines underwent some automatic post-processing before meshing; the small ending segments are cut out, the cycles are removed by computing a maximum spanning tree of the network, and the edges connecting the data points are re-oriented in the flow direction.

The meshes are created from centerlines using our meshing method and the method of \cite{abdellah2020interactive}, based on implicit structures. For the latter, we used the Blender plug-in provided by the authors, in metaballs reconstruction mode. Besides, we generated segmentation-based mesh from the segmented images using the Marching Cube algorithm. The mesh is smoothed using a Taubin filter and the disconnected components are removed to keep only the largest connected parts, following the common post-processing methods for segmentation-based meshing. In Section \ref{subsec:visual-eval}, we compare qualitatively the centerline-based and the segmentation-based meshes.
		
	\subsubsection{Datasets and segmentation models training}
	\label{subsec:comparison-datasets}
	
	We used data from two databases of healthy patients for this evaluation; 50 cerebral MRA images and 62 expert centerlines from the BraVa database, 34 cerebral MRA images from the TubeTK database \cite{bullitt2005vessel} and 34 in-house expert segmentations associated. These databases include the same type of images while offering complementary ground truth (respectively expert centerlines and expert segmentation). 
	
		To segment the vessels from the MRA images, we used two state-of-the-art methods based on neural networks; DeepVesselNet \cite{tetteh2020deepvesselnet} and U-net \cite{livne2019u}. We used the DeepVesselNet architecture provided by the author (https://github.com/giesekow/deepvesselnet) and we implemented the U-net neural network architecture. We trained the models on the expert segmentations of the TubeTK database; 27 images were included in the training set and 7 in the test set. 
	
	For the training, we used a combination of dice loss and cross-entropy loss as loss function and a stochastic gradient descent algorithm for the optimization, with a learning rate of 0.01 for U-net and 0.001 for DeepVesselNet. We set the batch size to 5 for U-net and 10 for DeepVesselNet due to memory constraints. These hyperparameters were set empirically by testing a large selection of values for each hyperparameter. We trained U-net for 200 epochs and DeepVesselNet for 300 epochs. Using the trained models, we segmented the vascular network for the 7 images of the test set of TubeTK and the 50 MRA of the BraVa database. Table \ref{tab:databases} summarizes to datasets used for the comparison study.
	
\begin{table}[h!]
\centering
\caption{Description of the different datasets used in our comparison study. The name of the original database (BraVa or TubeTK), the method used, the nature of the data, and the number of patients are given.}
\label{tab:databases}
\small
\begin{tabular}{|lllr|}
  \hline
Database & Method & Data & nPatients \\ 
  \hline
BraVa & DeepVesselNet & Segmentation & 50 \\ 
\rowcolor{Gray}
  BraVa & Expert & Centerlines & 62 \\ 
  BraVa & Unet & Segmentation & 50 \\ 
  TubeTK & DeepVesselNet & Segmentation & 7  \\ 
  TubeTK & Expert & Segmentation & 34  \\ 
  TubeTK & Unet & Segmentation & 7 \\ 
   \hline
\end{tabular}
\end{table}

\subsubsection{Quantitative evaluation}
\label{subsec:quantitative-eval}
	
In this section, we present the results of the quantitative evaluation of the segmentation-based centerlines compared to the expert centerlines according to the 9 topological and geometric features described hereafter. \textit{nBulges} corresponds to the number of bulges in the geometry, obtained by counting the ending segments smaller than the vessel diameter. \textit{nBranch} is the number of branches in the entire network. \textit{nCC} is the number of connected components and \textit{nBranchMaxCC} is the number of branches in the largest connected component. This metric highlights the extent of disconnected vessels and small isolated parts in the mesh. \textit{nCycle} is the number of cycles of the network. The only cycle in the cerebral vascular system is the circle of Willis, so the number of cycles should be either 1 for a complete circle of Willis or 0 for an incomplete circle of Willis. Finally, the branching topology of the network is analyzed via the number of bifurcations \textit{nBif}, the number of trifurcations or more \textit{nTrif+}, and the minimum (resp. maximum) furcation degree \textit{minDeg} (resp. \textit{maxDeg}), i.e. the number of in and out branches (bifurcations = 3). These metrics are reported in Table \ref{tab:topo-metric} for the different datasets considered.

\begin{table*}[ht]
\caption{topological and geometric features of the segmentation-based meshes for the different datasets. For each dataset, the median value between all patients is given.}
\label{tab:topo-metric}
\centering
\small
\begin{tabular}{|llrrrrrrrrr|}
  \hline
Database & Method & nBulges & nBranch & nCC & nBranchMaxCC & nCycle & nBif & nTrif+ & minDeg & maxDeg \\ 
  \hline
BraVa & DeepVesselNet & 43 & 369 & 451 & 59 & 26 & 76 & 15 & 0 & 5 \\ 
\rowcolor{Gray}
  BraVa & Expert & 0 & 205 & 1 & 205 & 0 & 102 & 1 & 1 & 4 \\ 
  BraVa & Unet & 38 & 504 & 380 & 200 & 44 & 136 & 30 & 0 & 6 \\ 
  TubeTK & DeepVesselNet & 52 & 552 & 557 & 212 & 46 & 140 & 28 & 0 & 6 \\ 
  TubeTK & Expert & 12 & 551 & 26 & 508 & 98 & 230 & 48 & 0 & 6 \\ 
  TubeTK & Unet & 34 & 626 & 300 & 446 & 85 & 215 & 46 & 0 & 6 \\ 
   \hline
\end{tabular}
\end{table*}

% Images to compare
\begin{figure*}[!h]
\centering
\includegraphics[width=0.9\textwidth]{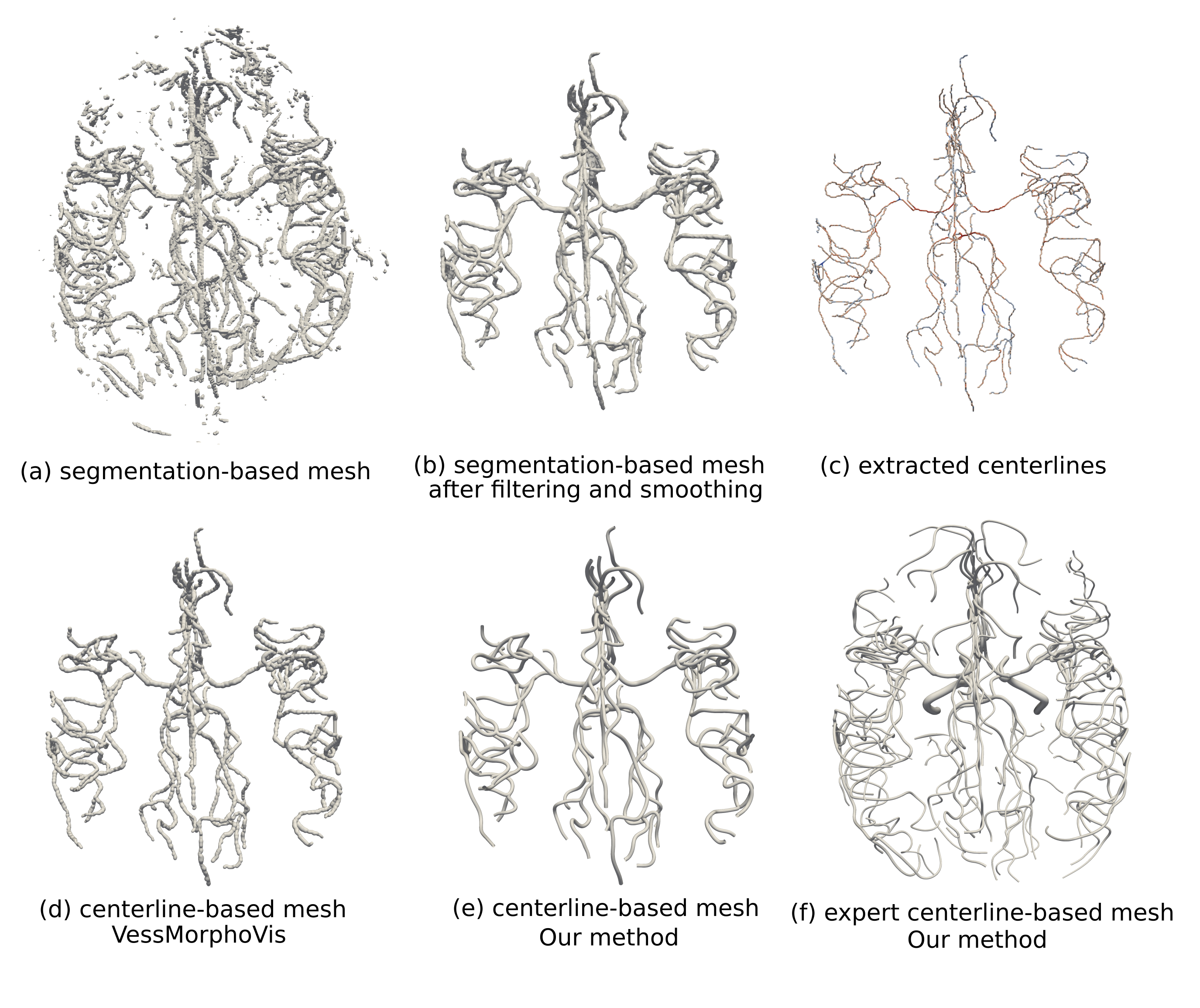}
\caption{Meshes produced with segmentation-based and centerline-based methods for a patient of the BraVa database. From left to right and top to bottom: original mesh created from the segmentation with Unet, the same mesh after filtering of the smallest components and smoothing; centerlines extracted from the mesh after post-processing; mesh produced by VessMorphoVis from these centerlines; mesh produced by our method from these centerlines; mesh produced by our method from manually extracted expert centerlines.}
\label{Img:comparison-mesh}
\end{figure*}

% Result analysis
 We observe in Table \ref{tab:topo-metric} that the expert centerlines (in gray) do not have any small ending segments ($nBulges = 0$). The network forms a single connected component, they are no isolated vessels and no cycles. Besides, the branchings are mainly bifurcations, as expected in the cerebral vascular system where trifurcations are rare. The expert centerlines show no branching with a degree superior to 4 (= trifurcations). 

On the other side, segmentation-based meshes present bulges ($> 12$) and cycles ($> 26$), mainly because closed vessels are merged in the resulting mesh. The number of trifurcations and higher degree branching is high ($> 15$), and furcations with up to 6 branches were observed. These metrics bring to light some inaccuracies in the topology of the meshes produced by segmentation, which will affect the mesh geometry and therefore the CFD simulation results. It is interesting to see that such problems (disconnected or merged vessels, bulges) are observed even in the meshes based on the ground truth segmentation made by medical doctors (see "TubeTK expert" row in Table \ref{tab:topo-metric}). It shows that errors are not only due to the performance of the segmentation method but also to the segmentation-based meshing process itself, as it relies on the segmentation of low-resolution images. Moreover, no distinction can be made between veins and arteries during segmentation, which might cause peculiar topology in the network. To run numerical simulations in such segmentation-based meshes, post-processing is required to isolate the arterial system and reconnect or separate vessels. The topological problems highlighted here are illustrated by enhanced visualizations of the meshes (Figures \ref{Img:comparison-mesh} and \ref{Img:comparison-zoom}) in the next section.

		\subsubsection{Visual evaluation}
		\label{subsec:visual-eval}

\begin{figure*}[!h]
\centering
\includegraphics[width=0.9\textwidth]{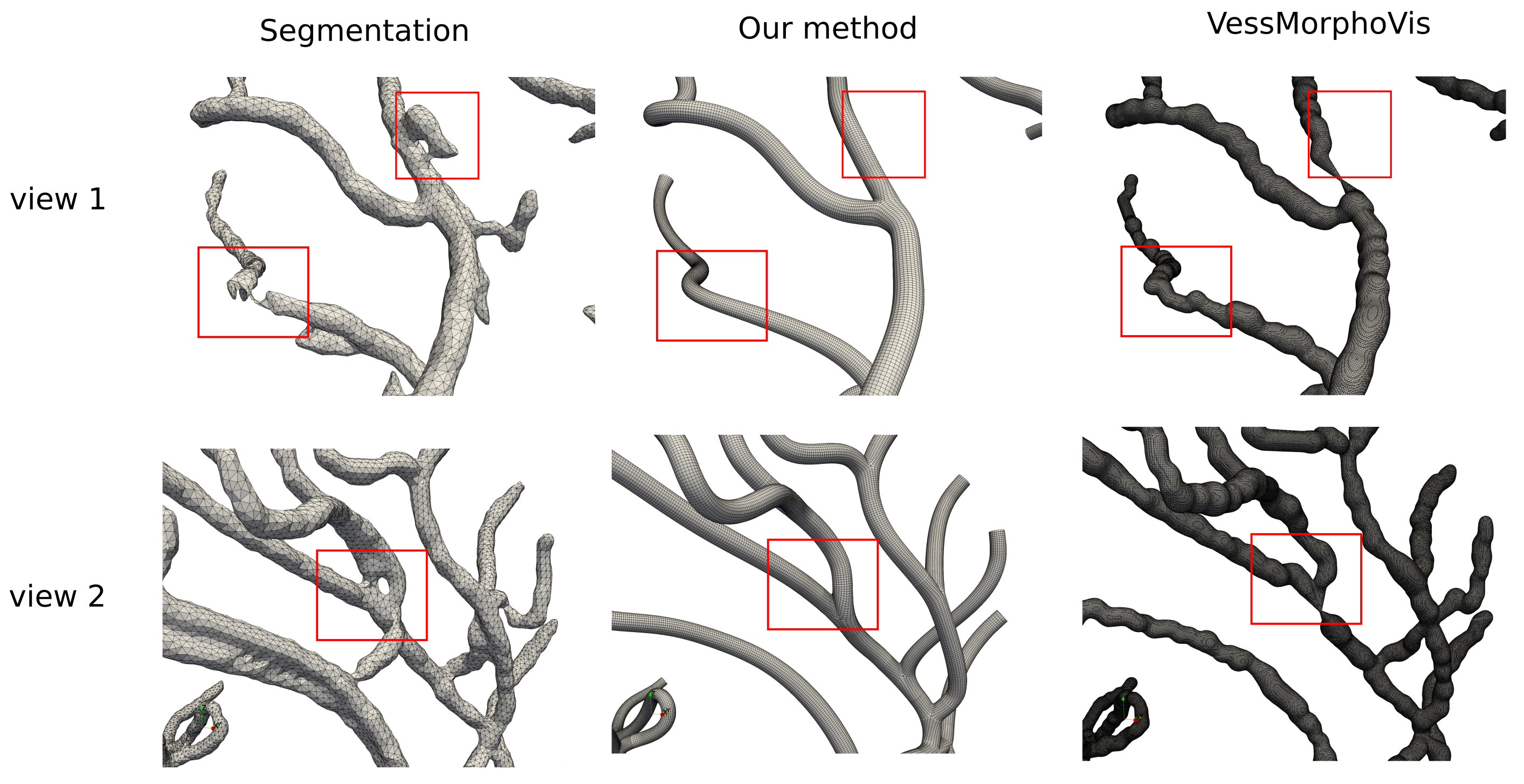}
\caption{Enhanced visualization of the segmentation-based mesh, the centerline-based mesh obtained with our method, and the centerline-based mesh obtained with VessMorphoVis. The red squares highlight the parts where the meshes show important differences.}
\label{Img:comparison-zoom}
\end{figure*}

		% Extraction centerlines from segmentations
		 In this part, the results from our method are visually compared against centerline-based meshes produced by the method of \cite{abdellah2020interactive} and the segmentation-based meshes produced by the method of \cite{tetteh2020deepvesselnet} and \cite{livne2019u}. We selected a BraVa patient for this visual evaluation to have access to the expert centerlines associated. We chose the segmentation produced by Unet for this patient as it provided better results than DeepVesselNet on this database. Figure \ref{Img:comparison-mesh} shows the whole brain meshes obtained by different methods, and Figure \ref{Img:comparison-zoom} shows enhanced visualization of some relevant parts. As shown in Figure \ref{Img:comparison-mesh} (a), the deep learning-based segmentation methods enabled to segment the whole brain vascular network, including small vessels. However, the mesh requires post-treatments such as the removal of the small isolated parts or smoothing (Fig. \ref{Img:comparison-mesh} (b)) to use for CFD. 
		  
	The centerline-based methods (images (d) and (e)) were able to produce meshes with a topology similar to the segmentation-based mesh from the centerlines automatically extracted from it. The mesh produced by our method is smoother than the other meshes for the same network. We discuss further the geometric quality of the meshes in the analysis of Figure \ref{Img:comparison-zoom}. As illustrated in Image (f), the manually extracted expert centerline led to a larger arterial network with smaller vessels than the segmentation. The radius of the expert centerlines is smaller, due to the extraction method \cite{longair2011simple}.
		
	 As shown in View 1 of Figure \ref{Img:comparison-zoom}, our method, which relies on the vessel tubularity assumption, cleared the vascular network from the bulges observed in the segmentation-based mesh. The radius and trajectory smoothing allows for reconstructing the disconnected parts. View 2 highlights the merging vessels and cycles observed in segmentation-based meshes. The automatic post-treatment of the centerlines allowed us to remove unwanted cycles in the network.

The implicit method VessMorphoVis \cite{abdellah2020interactive} enables to mesh complex branching patterns. However, as shown in Figure \ref{Img:comparison-zoom}, this method is sensitive to noise on the centerline geometry and radius. This cause the surface of the vessels to appear bumpy and irregular, which will impact the CFD simulations. Our method improves the smoothness of the vascular geometry. Those results demonstrate the ability of our algorithm to produce meshes not only from manually extracted centerlines but also to integrate fully-automated pipelines.

	\subsection{Mesh quality}
	\label{subsec:Meshquality}
	
	\begin{figure}[ht]
\centering
\includegraphics[width=0.45\textwidth]{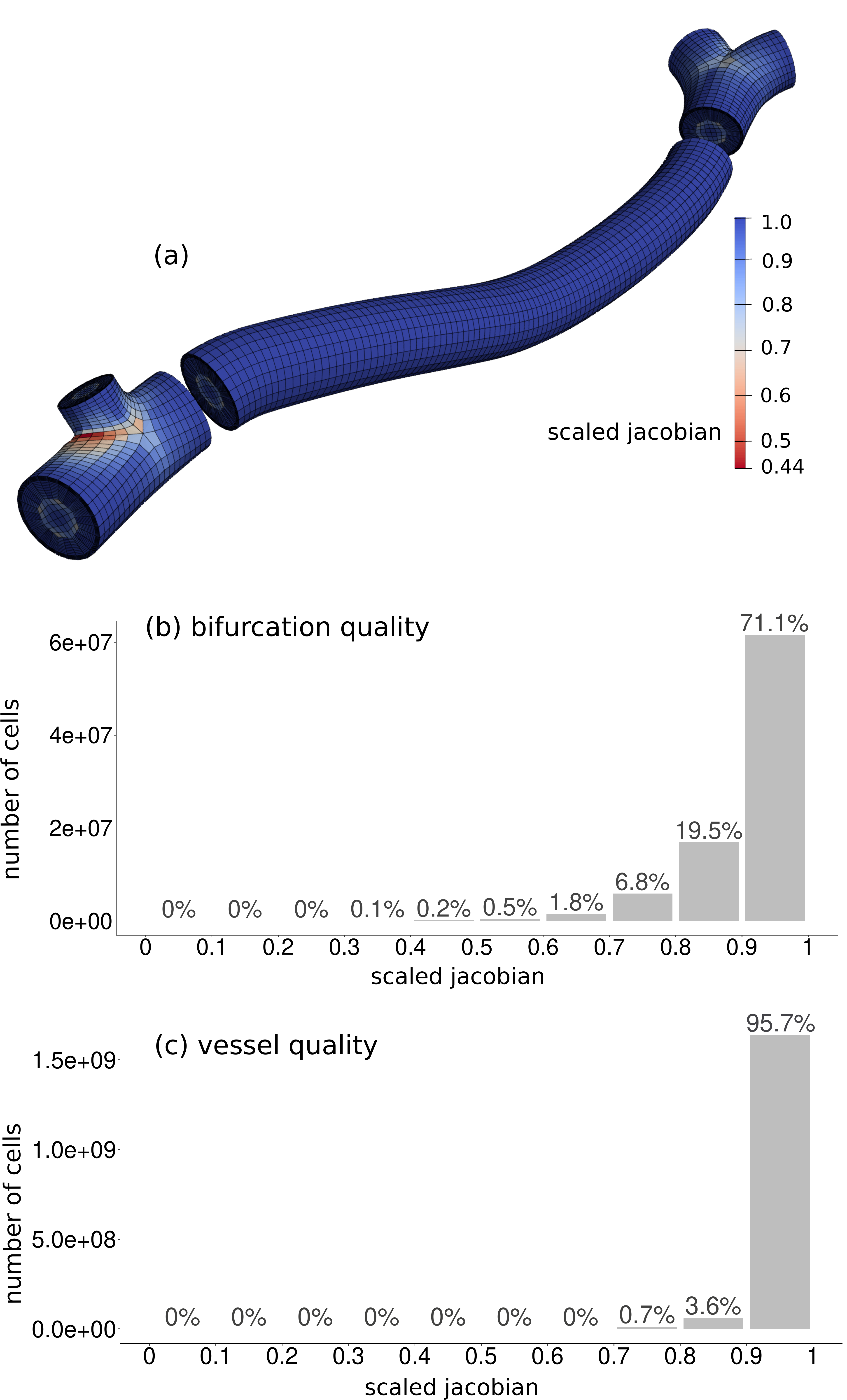}
\caption{Distribution of the scaled Jacobian values of the mesh cells. Histogram (b) represents the quality of bifurcation cells and histogram (c) the quality of the vessel cells. Image (a) illustrates the location of high and low-quality cells within a mesh.}
\label{Img:histograms}
\end{figure}

	In CFD, the accuracy and stability of the simulation are affected by the mesh quality. We assessed the quality of the meshes generated with the proposed method through the scaled Jacobian of the cells. The scaled Jacobian ranges from -1 (worst quality) and 1 (best quality). Negative values indicate invalid cells. The volume meshes for 60 patients from the BraVa database were generated  (see section \ref{subsec:Largecerebralarterialnetwork} for details), with the following parameters; $N = 24$, $d = 0.2$, $\alpha = 0.2$, $\beta = 0.3$ , $\gamma = 0.5$, $N_{\alpha} = 10$, $N_{\beta} = 10$. We evaluate the cells of the bifurcations and vessels separately. We excluded from the study the bifurcations and vessels for which the meshing failed (scaled Jacobian $<0$), as explained in Section \ref{subsec:Largecerebralarterialnetwork}. The histograms of scaled Jacobian for the 60 patients are given in Figure \ref{Img:histograms}.
	
	As shown in Image (a) of Figure \ref{Img:histograms}, the low-quality cells are mainly localized in the bifurcation separation planes. Nevertheless, we achieved a very good quality for bifurcation cells, with $71\%$ of the cells with a scaled Jacobian value higher than $0.9$. The vessel cells have even better quality, with $95.7\%$ of the cells having a scaled Jacobian higher than $0.9$. In terms of mesh quality, our method improves the state of the art. Indeed, only $49\%$ of the cells have a scaled Jacobian above $0.9$ on average on the distributions given for three large cerebral networks in \cite{ghaffari2017}. This proportion goes up to $62\%$ of the cells for the abdominal aortic artery geometry meshed by the method of \cite{xiong2013automated}. Finally, in \cite{deSantis2011}, between $65\%$ and $82\%$ of the cells of the aortic arch meshed have a scaled Jacobian value between \textbf{$0.8$} and $1$. Quantitatively, our method gives better results. However, we bear in mind that the study of \cite{deSantis2011} and \cite{xiong2013automated} focuses on arterial geometries that differ from our study and that the histogram of cell quality depends on the mesh cell density.
	
	\subsection{Computation time}
	\label{subsec:Computationtime}

	We measured the computational time of the modeling and meshing steps for three patients of the BraVa database. The results are given in Table \ref{Tab:time}. The average time for modeling a large cerebral vascular network is about 16 minutes. We measured the meshing time for different cell densities. The average meshing time goes from 24.6 minutes for a coarse mesh to 49.7 minutes for a fine mesh. We want to stress that this study was performed on large networks, with a high number of bifurcations (around $100$) and vessels (around $200$). The meshing time increases with the number of bifurcations and vessels, while the modeling time is affected by the number of data points. 
	
\begin{table}[!h]
\caption{Computational time required to model and mesh large vascular networks from the BraVa dataset.\\}
\label{Tab:time}
\centering
\addtolength{\tabcolsep}{-4pt}
\begin{tabular}{|cccccc|}
  \hline
  \begin{tabular}{c} furcation \\ (\#) \end{tabular} &  \begin{tabular}{c} vessel \\ (\#) \end{tabular}  & \begin{tabular}{c} data point \\ (\#) \end{tabular} & \begin{tabular}{c}  modeling \\  time (min) \end{tabular} &  \begin{tabular}{c} cells \\ (\#) \end{tabular} &  \begin{tabular}{c}  meshing \\ time (min) \end{tabular} \\ 
\hline
 96 & 194 & 2816 & 11.3 &  \begin{tabular}{r} 1389k \\ 1853k \\ 2316k \\ 2779k \end{tabular} & \begin{tabular}{r} 20.4 \\ 25.7 \\ 31.2 \\ 38.4 \end{tabular}\\
\hline
101 & 203 & 3531 & 18.3  & \begin{tabular}{r} 1916k  \\ 2555k \\ 3193k \\ 3832k \end{tabular} &  \begin{tabular}{r} 27.5 \\ 38.2 \\ 49.1 \\ 67.5 \end{tabular} \\
\hline
 107 & 216 & 3474 & 16.8 & \begin{tabular}{r} 1737k \\2316k \\ 2895k \\3474k  \end{tabular} & \begin{tabular}{r} 26.3 \\ 36.1 \\ 44.2 \\ 55.9 \end{tabular} \\
\hline
\end{tabular}
\end{table}

Besides, a large part of the meshing time corresponds to the computation of the surface nodes; on average 17.4 minutes for a coarse mesh and 34.8 minutes for a fine mesh. The volume mesh is generated directly from the nodes of the surface mesh without recomputing them. Finally, meshing can be run in parallel, by splitting the network into parts to be meshed on different CPUs. Using 12 CPUs, we reduced the meshing computational times given in Table \ref{Tab:time} by a factor of 5.

\section{Applications}
\label{sec:Applications}

In this section, we propose several applications of the proposed modeling and meshing framework.

	\subsection{Deformation}
	\label{subsec:Deformation}

\begin{figure}[ht]
\centering
\includegraphics[width=0.5\textwidth]{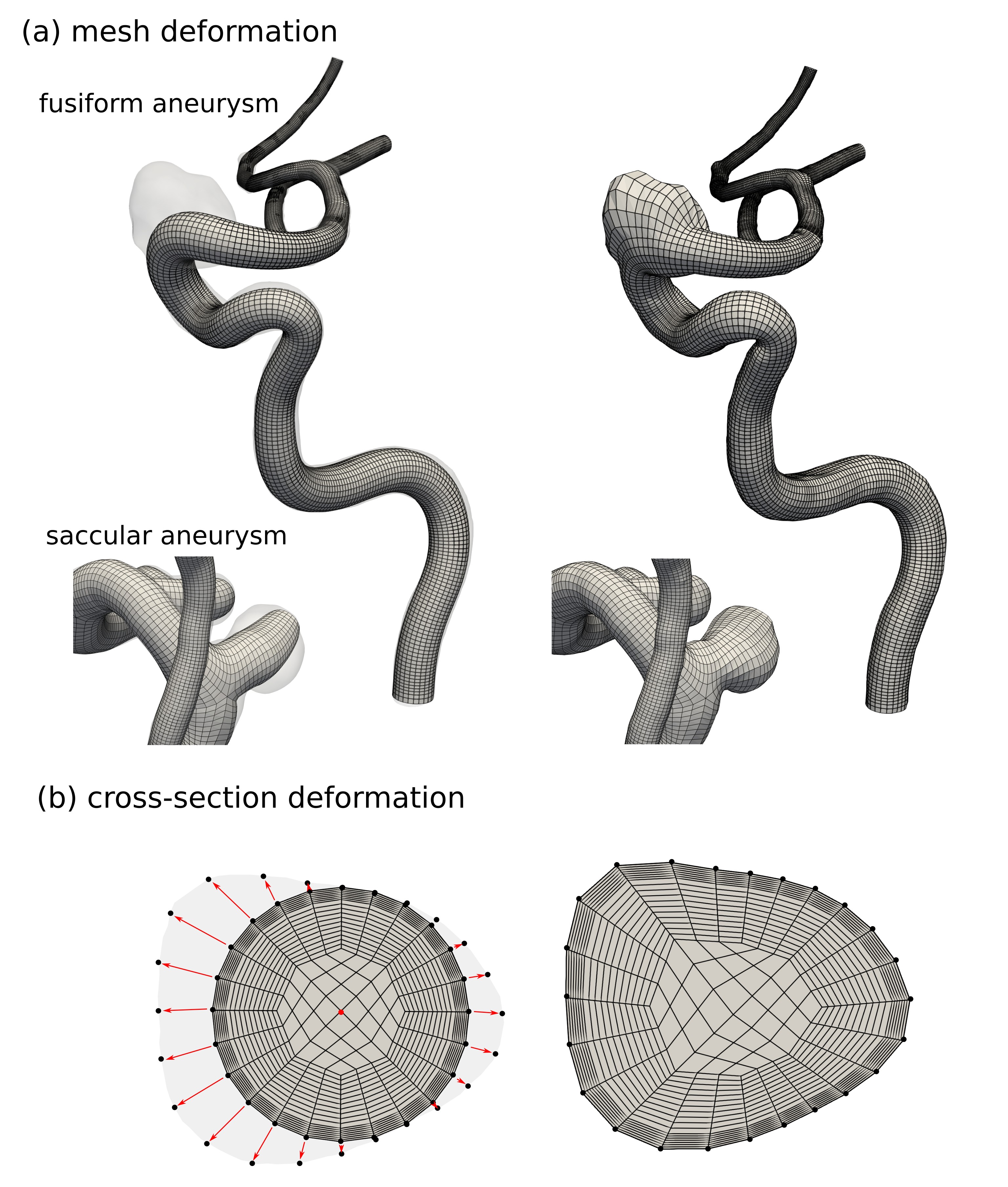}
\caption{(a) Structured hexahedral meshing of cerebral arteries with a fusiform or saccular aneurysm by deformation. On the left, the tubular mesh is superimposed on the target surface. On the right, we show the mesh after projection. (b) Cross-section pattern before and after deformation.}
\label{Img:remeshing}
\end{figure}

	The proposed model relies on the assumption that vessel cross-sections are circular, which is limiting when dealing with pathological vessels. A way to address this limitation is to deform the cross-sections to match a target surface as post-processing. If the user input data is a surface mesh, we propose the following alternative use of our meshing framework:

	\begin{enumerate}
	\item{Extract the centerline from the surface mesh (VMTK),}
	\item{Create a tubular mesh from the centerline using the proposed method,}
	\item{Deform the tubular mesh to match the original surface.}
	\end{enumerate}
	
Figure \ref{Img:remeshing} illustrates an example of this pipeline to mesh arteries with aneurysms. In the deformation step, we project the nodes onto the surface of the target mesh, radially from the cross-section center. Saccular aneurysms are initially modeled as bifurcating vessels and then deformed. Because the shape of the volume mesh pattern depends on the position of the section nodes (cf Section \ref{subsec:ModelVessels}), the deformation of the surface mesh is smoothly conveyed to the cells inside the mesh, as shown in Figure \ref{Img:remeshing} (b). We can use this pipeline to re-mesh a triangular surface mesh with hexahedral cells. However, important deformations may impact the quality of the cells and cause cells to intersect in high-curvature areas.

	\subsection{Topology and geometry editing}
	\label{subsec:Topologyediting}
	
	 The relationship between the vascular tree topology and geometry (e.g. the different configuration of the circle of Willis, vessel angle) and the hemodynamics have been studied extensively in the literature, using ideal or patient-specific models \cite{cornelissen2018aneurysmal,alnaes2007computation}. The proposed meshing framework finds applications in creating and editing vascular models. As our modeling method requires only a few data points, the centerlines can be easily created or edited to modify the bifurcation angles, the radius, or the trajectory of a vessel.
	 
\begin{figure}[!h]
\centering
\includegraphics[width=0.5\textwidth]{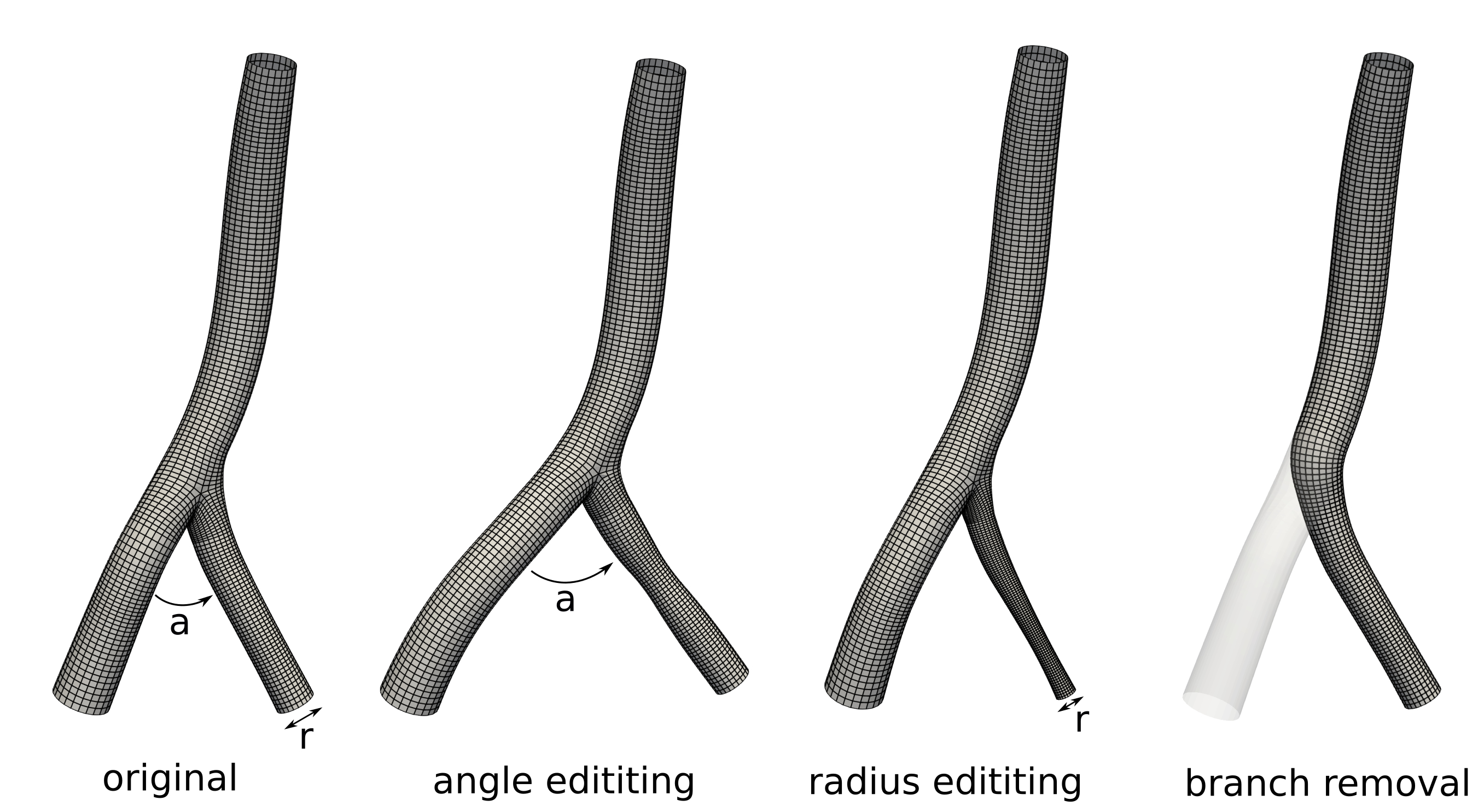}
\caption{Editing of a model of the basilar artery. We modified the bifurcation angle and the radius of the original vertebral artery and removed one of the vertebral arteries.}
\label{Img:edit_geom}
\end{figure}
	 
	 Figure \ref{Img:edit_geom} provides examples of such modifications. The graph structure encoding proposed (Section \ref{subsec:Vessels}) facilitates the identification and modification of the data points of a branch of interest. We can adjust the model parameters (bifurcation cross-sections and apex smoothing, vessel smoothing) and mesh (cell longitudinal and circumferential density, boundary layer). As the bifurcations are modeled by two merging vessels, one branch can be removed without affecting the trajectory of the other branch, as illustrated on the right in Figure \ref{Img:edit_geom}. Those modifications are performed without additional computational cost by a local re-computation of the model and mesh parts.

	\subsection{Large cerebral arterial network meshing}
	\label{subsec:Largecerebralarterialnetwork}
	
\begin{figure}[!ht]
\centering
\includegraphics[width=0.45\textwidth]{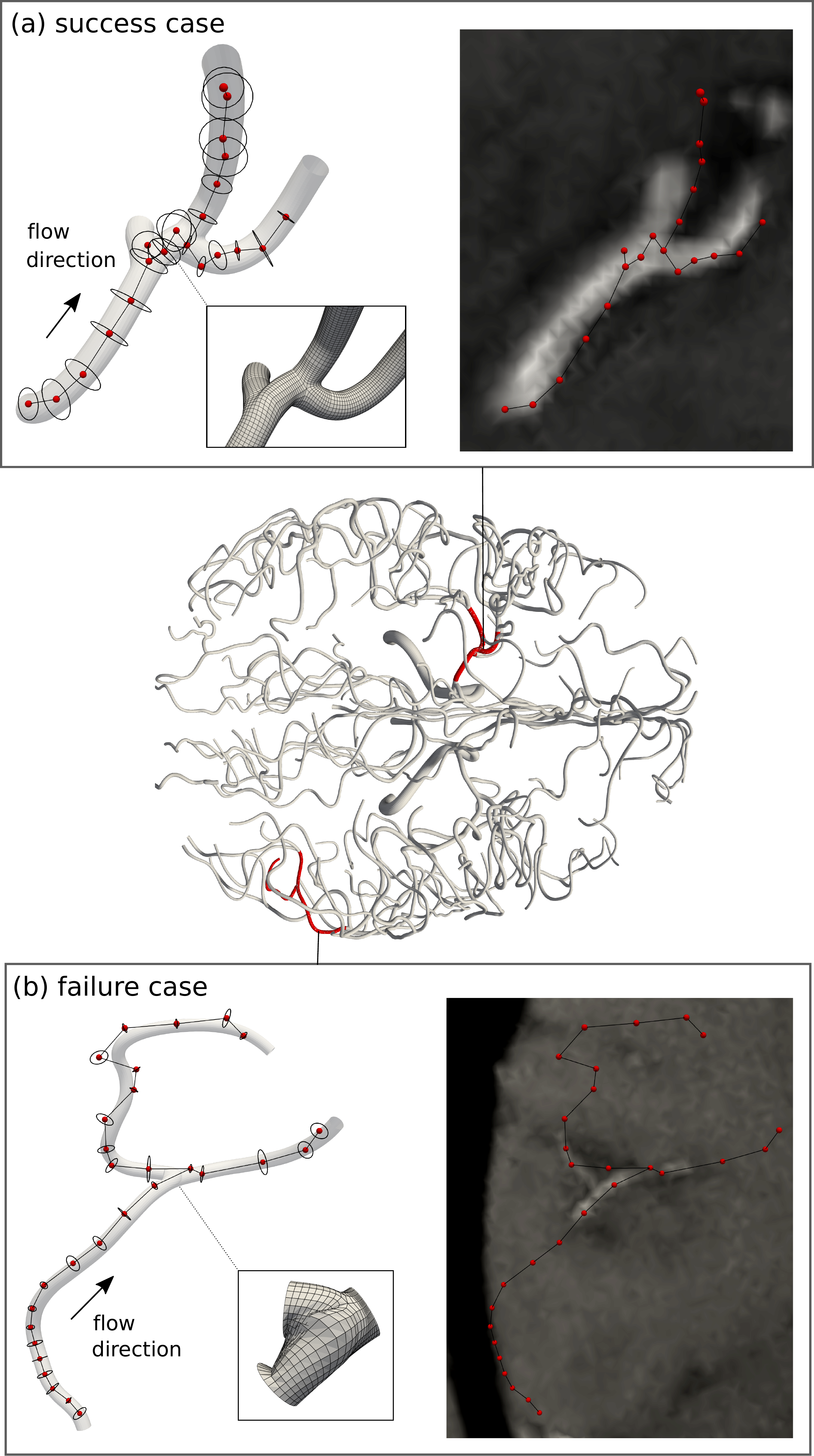}
\caption{Example of success and failure of our method for one patient of the BraVa database. The whole-brain mesh is represented in the middle with a focus on two parts of the network. For each focus, the original centerline data points are represented by red dots (center) and black circles (radius). The mesh obtained is superimposed on the data points, with a highlight on the relevant parts. On the right image, the original centerline data points are overlayed on the original MRA image.}
\label{Img:fail-case}
\end{figure}

\begin{figure*}[!ht]
\centering
\includegraphics[width=1\textwidth]{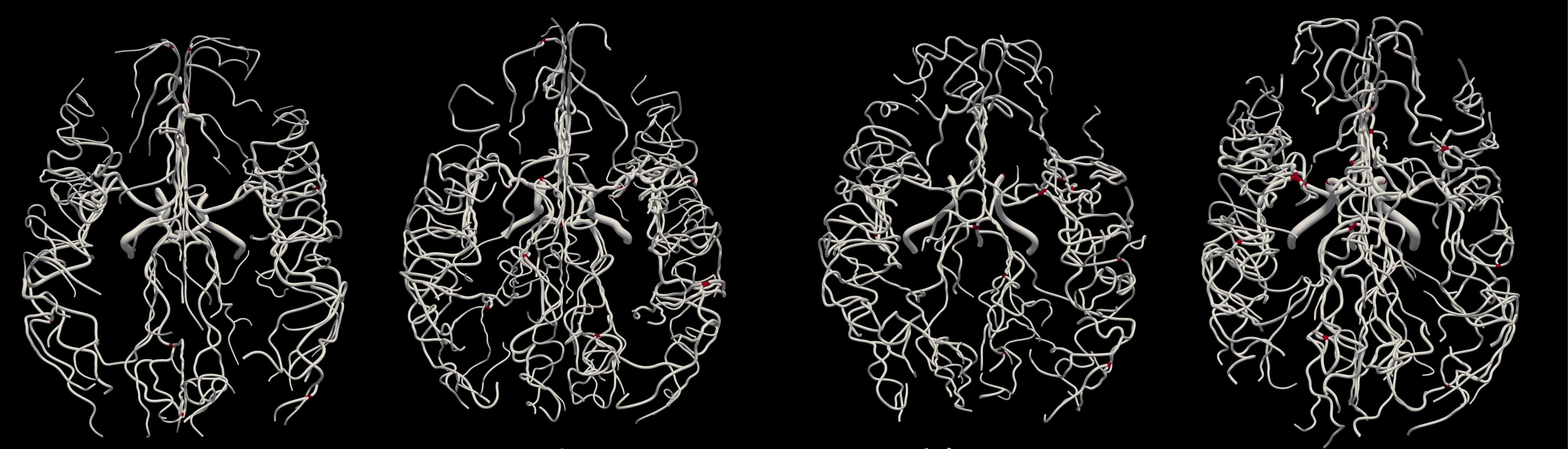}
\caption{Top view of 4 meshes among the 60 generated from the patients of the BraVa database. The bifurcations where the meshing algorithm has failed (i.e. at least one of the cells has a negative Jacobian) are represented in red. The cross-sections of the vessels with cells of negative Jacobian values are also represented in red.}
\label{Img:brava}
\end{figure*}
	
One of the objectives of the proposed method was to create meshes for CFD from existing databases of centerlines. Using our method, we created meshes for 60 patients of the BraVa dataset. These meshes are shown for 4 patients in Figure \ref{Img:brava}. The BraVa dataset is considered challenging for several reasons. The centerlines were semi-automatically extracted by medical doctors, and therefore suffer from limitations such as low sampling or data point misplacement. The superimposition of the centerline data points on the magnetic resonance angiography image in Figure \ref{Img:fail-case} shows the high level of noise encountered in the centerlines, both in the radius estimation and the spatial positions. Besides, by computing the ratio of the number of data points on the total length of the connecting polyline, we estimated the average point density in the database to $0.45$ $mm^{-1}$, which is very low. 
 
In this database, we evaluated the percentage of successfully meshed vessels and bifurcations. We consider that the meshing has failed if at least one cell has a negative scaled Jacobian score. With this strict definition, $83\%$ of the bifurcations and $92\%$ of the vessels were successfully meshed. 
 
The reason for the failure of the vessel mesh is a too-high curvature occurring mainly in the arteries with high tortuosity such as the internal carotid arteries. The causes of failure for the bifurcations are small branching angles and misplaced data points at the branching parts. This last case is illustrated in Image (b) of Figure \ref{Img:fail-case}. We can see that the bifurcation point in the centerline data was positioned downstream in the main vessel, causing one of the daughter vessels to go backward from the direction of the flow with a sharp angle. It led to a wrong estimation of the parameters of the bifurcation model, which did not represent the bifurcation shape correctly and caused the meshing to fail. Figure \ref{Img:fail-case} (a), on the other hand, illustrates a successful reconstruction. Although the input centerline was very imprecise both in the radius estimation and point positions, we were able to create a smooth mesh close to the geometry given by the medical image, demonstrating the advantages of our vessel and bifurcation models. As shown in the insert of Figure \ref{Img:fail-case} (a), challenging topologies with short connecting segments between bifurcations were successfully meshed with hexahedral elements. An image of all the meshes of the database, with failure areas highlighted, is given in Supplementary Materials, section 1.3.

\subsection{CFD simulation}
\label{subsec:CFDsimulation}

\begin{figure}[!ht]
\centering
\includegraphics[width=0.5\textwidth]{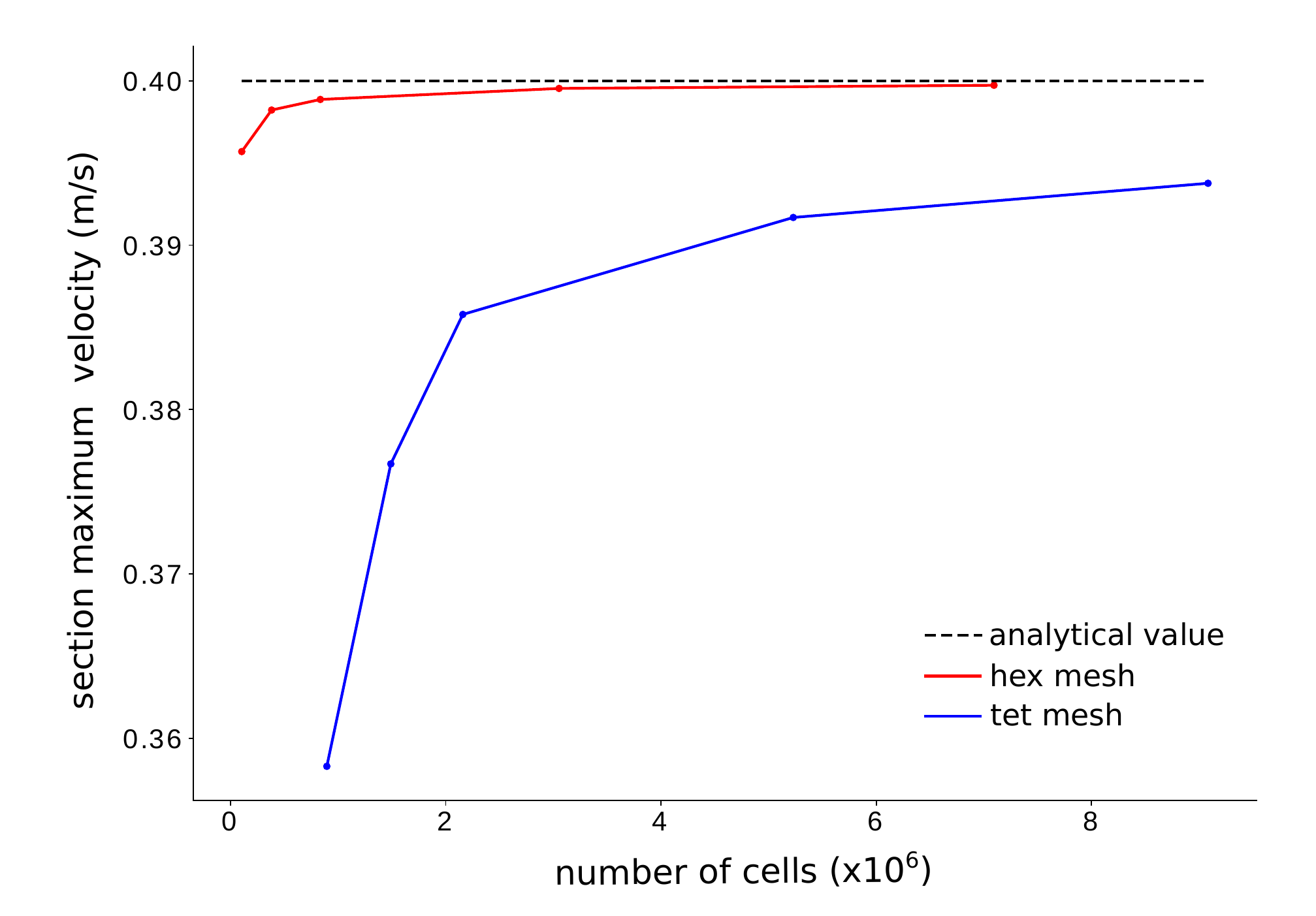}
\caption{Graph of the sectional maximum velocity as a function of the number of cells in the mesh for both tetrahedral meshes and hexahedral meshes. The maximum velocity was averaged on three cross-sections along the tube model. The analytical value expected is shown by the black dotted line. }
\label{Img:cfd-tube}
\end{figure}

\begin{figure*}[!ht]
\centering
\includegraphics[width=0.7\textwidth]{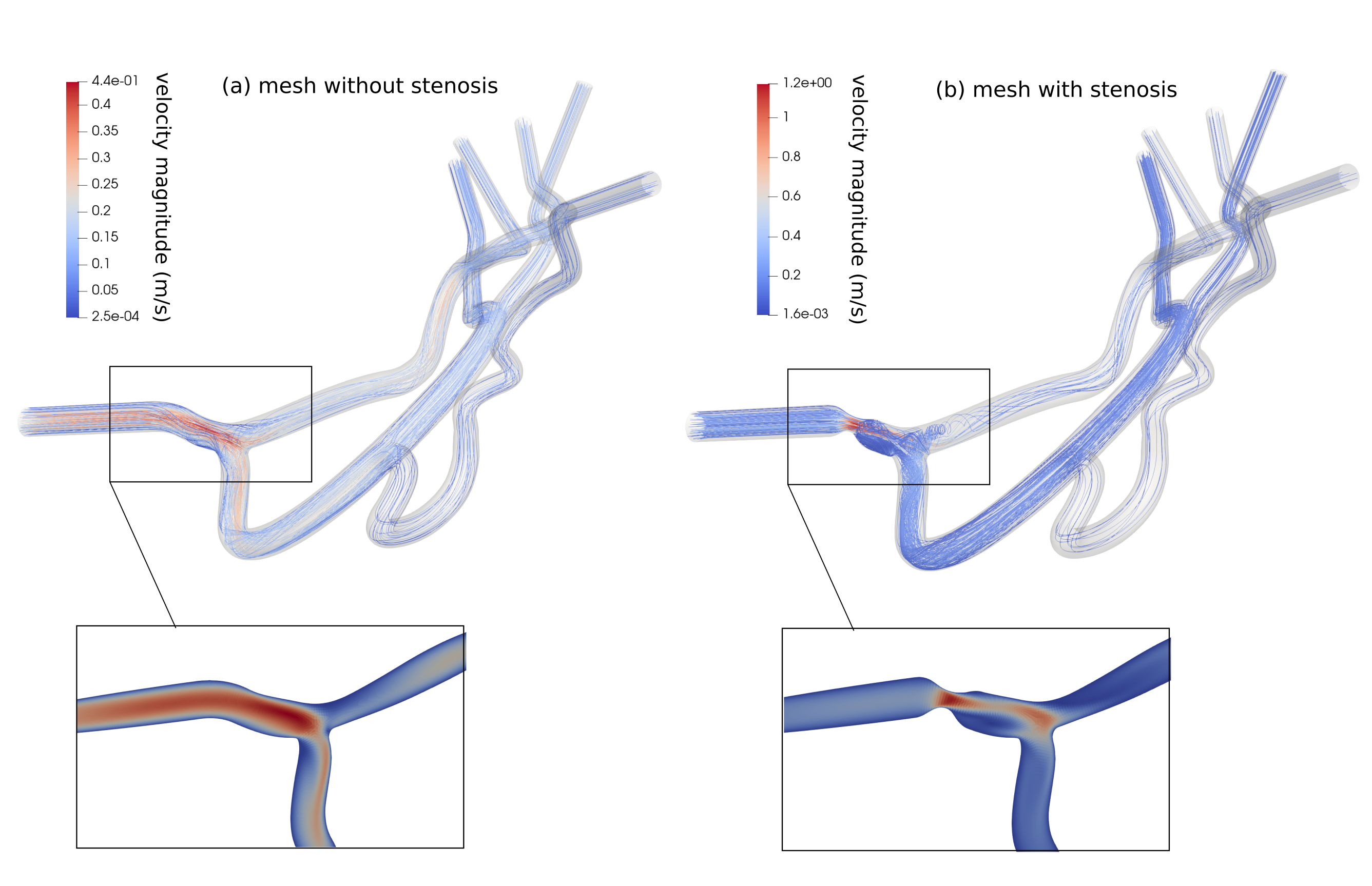}
\caption{CFD simulation results for the mesh without stenosis (a) and with stenosis (b). In both cases, the velocity streamlines were rendered and the velocity field is shown by a cut on the stenosis region.}
\label{Img:cfd-velocity}
\end{figure*}

In this section, we demonstrate the applicability of the meshing method proposed for CFD simulations. Firstly, we compare the hexahedral meshes produced by our method to the commonly used tetrahedral meshes in terms of computational cost, convergence, and accuracy of the results. This comparison was conducted in a straight tube model, as this experimental setting allows a comparison to the analytical form of the sectional velocity profile given by the Poiseuille equation. We set the tube diameter to mimic a middle cerebral carotid artery ($D = 2.5 mm$), and the tube length to guarantee that the flow is fully developed ($L= 200 mm$). We generated five tetrahedral volume meshes of increasing cell density (from coarse to fine)using the software TetGen\textregistered \ \cite{hang2015tetgen}, a state-of-the-art tetrahedral meshing software often used to produce the volume mesh in blood flow studies \cite{taebi2020computational, shad2021patient}. In the same way, we generated five hexahedral volume meshes of increasing cell density with our method. For the CFD simulations, we selected the fluid properties to mimic blood, with a density $\rho = 1053 kg.m^{-3}$, and a dynamic viscosity $\mu = 0.0035 kg.m^{-1}.s^{-1}$. The flow was assumed laminar, as justified by a Reynold number of $150.4$. We set the inlet boundary condition to a fixed velocity $U = 0.2m.s^{-1}$, and the outlet boundary condition to zero pressure. We set the residual value for convergence to $10^{-6}$. We performed the CFD simulations using ANSYS Fluent (ANSYS Inc., USA).

As shown in Figure \ref{Img:cfd-tube}, the mesh independence was reached faster using hexahedral meshes than tetrahedral meshes, for a more accurate sectional maximum velocity value. The convergence of the simulation was also improved, as 4 times fewer iterations were necessary to obtain convergence of the results with hexahedral meshes. The simulation time was reduced on average by a factor of 3, which adds to the fact that fewer cells are required to reach accurate results with hexahedral meshes (as shown in Figure \ref{Img:cfd-tube}), reducing the computational cost even more. These results are consistent with the conclusions given in the works of \cite{vinchurkar2008evaluation}, \cite{de2010patient}, and  \cite{ghaffari2017}, demonstrating the advantages of hexahedral meshes over tetrahedral meshes for CFD simulations. We provide additional information on the methods and results in Supplementary Materials, Section 3.2.1. We reproduced this experiment in a realistic bifurcation model, leading to similar results, reported in Section 3.2.2 of the Supplementary Materials.

Secondly, we applied our method to reconstruct a patient-specific mesh of the middle carotid artery (MCA) and downstream vessel in a case where the segmentation failed to produce a valid mesh. We used the proposed framework to automatically extend the mesh at the inlet and outlet and add a stenosis on the MCA. The stenosis was designed to induce a reduction of $50 \%$ of the vessel diameter. We provide some images of the meshes in Supplementary Materials, Section 3.3.  

We used simple boundary conditions, as our goal is not to provide an analysis of this case study but to demonstrate the applicability of our method for the study of cerebrovascular pathologies by CFD. Blood is considered a Newtonian fluid ($\rho = 1053 kg.m^{-3}$, $\mu = 0.0035 kg.m^{-1}.s^{-1}$), and the flow is assumed steady and laminar. We set the inlet velocity to $0.2m.s^{-1}$ \cite{lindegaard1987variations}, and the outlet pressure to zero. The simulation converged with a residual value of $10^{-6}$ in 50 iterations and 78 iterations, in 4 and 6 minutes for the healthy and pathological cases respectively. The velocity streamlines and the velocity fields computed by CFD for the healthy and pathologic cases are shown in Figure \ref{Img:cfd-velocity}. With this experiment, we demonstrated the potential of our meshing method to easily design and conduct blood flow studies by CFD. The editing flexibility of our framework allows us to study the effect of hemodynamic pathologies or topological changes compared to a reference geometry like in Figure \ref{Img:cfd-velocity}. The advantages of our framework are not limited to modeling and meshing, as it also facilitates the analysis of the results (e.g. extraction of cross sections, extraction of velocity values along the centerline). 

\section{Conclusion}

In this article, we addressed the problem of the reconstruction and meshing of large vascular networks from centerlines. We put into light the gap existing between segmentation and CFD simulation. We proposed a method to fill this gap using centerlines. This method helps to overcome the shortcomings of segmentation-based meshes and facilitates the creation and editing of meshes of large vascular networks suitable for CFD. It is robust to the low sampling and noise of the input centerlines, which makes it compatible with automatically and manually extracted centerlines. It allows us to take advantage of the existing databases. The hexahedral meshing proposed improves the accuracy and reduces the cost of numerical simulations compared to commonly used tetrahedral mesh, which can have an impact on clinical studies \cite{lewandowska2019meshing}.

We acknowledge some limitations to this work. We originally developed this meshing framework for cerebral vascular networks. We have not addressed yet the meshing of non-planar n-furcations ($n>3$) that are common in other vessels (e.g. aorta, lung vessels), hindering the generalization of our method to any vascular networks. In addition, the robustness of the modeling method needs to be further improved as it causes the meshing to fail in some cases. For this, we would like to integrate more anatomical constraints on the bifurcation and vessel models (maximum curvature, maximum bifurcation angle). Besides, we want to emphasize that our objective with this work was not to improve the performance of the segmentation or centerline extraction algorithms but to acknowledge the limitations of the existing data and generate realistic meshes suitable for CFD from flawed centerlines and existing databases. Hence, the accuracy of the reconstruction depends on the quality of the input centerlines; some manual post-treatment may still be required before simulation. In this way, our framework offers more editing flexibility than other meshing methods. To take advantage of this flexibility, we developed a vascular network editing software, with a user-friendly interface \cite{decroocq2022software}. This interface integrates the modeling and meshing methods described in this article and other editing functionalities (e.g. centerline editing, branch removal, or angle modification). It is well suited for the creation of meshes for the study of the impact of topology and geometry of the vascular network on blood flow. It opens vascular modeling and hexahedral meshing to medical doctors and non-expert users.

\section*{Acknowledgments}

We would like to thank Simon Tupin, who initiated this work and Erwan Maury for his contribution of the development apex smoothing method. Founding is acknowledged through AURA region (SIMAVC project) and ANR 20-CE45-0011 (PreSpin project).

\bibliographystyle{plain}
\bibliography{refs}

\end{document}